\documentclass{article}
\usepackage{amssymb}
\usepackage{graphicx}

\title
{\bf Stretching and relaxation dynamics in double stranded DNA}
\author{{\bf D. Hennig$^1$\footnote{Corresponding author:
Email:hennigd@physik.fu-berlin.de}
and J.F.R. Archilla$^{2}$}
\\
\\
Freie Universit\"{a}t Berlin, Fachbereich Physik
\\
Institut f\"{u}r Theoretische Physik
\\
Arnimallee 14, 14195 Berlin, Germany\\
\\
\\
$^2$ Group of Nonlinear Physics\\ Departamento de F\'{i}sica
Aplicada I\\ ETSI Inform\'{a}tica, University of Sevilla,\\ Avda
Reina Mercedes, s/n. 41012\,-\,Sevilla, Spain\\}

\begin{document}
\newcommand{\singlefig}{.75\textwidth}
 \newcommand{\doublefig}{\textwidth}
 \newcommand{\tripfig}{0.65\textwidth} 

\maketitle


\begin{abstract}
\noindent  We study numerically the mechanical stability and
elasticity properties of duplex DNA molecules within the frame of
a network model incorporating  microscopic degrees of freedom
related with the arrangement of the base pairs. We pay special
attention to the opening-closing dynamics of double-stranded DNA
molecules which are forced into
 non-equilibrium conformations. Mechanical stress imposed at one terminal
end of the DNA molecule brings it into a partially opened
configuration. We examine the subsequent relaxation dynamics
connected with energy exchange processes between the various
degrees of freedom and structural rearrangements leading to
complete recombination to the double-stranded conformation. The
similarities and differences between the relaxation dynamics for a
planar ladder-like DNA molecule and a twisted one are discussed in
detail. In this way we show that the attainment of a
quasi-equilibrium regime proceeds faster in the case of the
twisted DNA form than for its thus less flexible ladder
counterpart. Furthermore we find that the velocity of the complete
recombination of the DNA molecule is lower than the velocity
imposed by the forcing unit which is in compliance with the
experimental observations for the opening-closing cycle of DNA
molecules.
\end{abstract}

\newpage

\section{Introduction}
Nowadays powerful micromanipulation techniques allow for the
selective excitation of DNA in single molecule experiments under
controlled conditions (see, e.g.,
\cite{Essevaz},\cite{Bockelmann1},\cite{Clausen}). Particularly
mechanical properties of DNA molecules under tension and the
inducement of structural transitions with external stress have
received a lot of attention \cite{Marko}-\cite{Nelson}. In this
context several force measurements were performed on single DNA
molecules to examine their elastic response to applied stretching
\cite{Smith} and twisting forces, respectively
\cite{Marko},\cite{Strick},\cite{Nelson}. It has been demonstrated
that with the application of stronger forces to the molecule in
stretching experiments, even highly cooperative conformational
transitions were discovered, for which, e.g., B-DNA is converted
into a new overstretched conformation called S-DNA \cite{Smith}.
Furthermore, with sufficiently high tension the mechanical
unzipping of single DNA molecules is achievable \cite{Bockelmann}.
The opening process of DNA has been the subject of a number of
studies due to its immense relevance for the transcription
process. Simple models taking into account the elasticity
properties of DNA, such as the wormlike and rodlike chain model,
respectively have been invoked to describe these unzipping
experiments \cite{elastic}. The statistical mechanics of the
torque induced denaturation of DNA has been studied in
\cite{Cocco99}. Recently, a semi-microscopic model of the binding
between the two strands has been used to explain the effects of
applied tension to DNA unzipping \cite{Cocco02}. In order to
determine the geometry and deformability of DNA at a local level,
i.e. in terms of base pairs, energy functions based on discrete
network models have been derived in \cite{Gorin}-\cite{Coleman}.
In an early network model for the base pair opening in DNA
involving microscopic degrees of freedom, Peyrard and Bishop (PB)
proposed a planar ladder-like oscillator model of DNA assigning
each base pair a vertical inter-strand vibrational degree of
freedom which simulates the stretchings and compressings of the
corresponding H-bridges \cite{PB}. The binding forces of the
hydrogen bridges are described by a Morse potential. The bases
itself are treated as point masses. Horizontally, the bases on the
same strand are coupled via the stacking interaction described by
harmonic potentials. The PB model has been extensively studied and
localized oscillating solutions (breathers) have been found
reflecting successfully some typical features of the DNA opening
dynamics such as the magnitude of the amplitudes and the time
scale of the breathing of the 'bubble' occurring prior to thermal
denaturation \cite{PB}. On the other hand, as the coiled form of
double-stranded DNA is concerned, it has been found that the
bubble formation is strongly correlated with twist deformations
and local openings are always connected with a local untwist of
the double helix \cite{BCP},\cite{Barbithesis}. In order to
account for the helicoidal structure of DNA the PB model has been
significantly extended by Barbi, Cocco and Peyrard (BCP)
\cite{BCP}. In their model of the twisted form of DNA (the
Watson-Crick double helix), two degrees of freedom per base pair
are introduced. There is a radial variable measuring the distance
between  two H-bridged bases along a line that connects them in
the base plane being perpendicular to the helix axis. Further, the
twist angle between this connecting line and a reference direction
determines the orientation of the H-bridge. In this model a
further degree of freedom has been included, namely the axial
distance variation \cite{Cocco02}. Interactions between the bases
are described then by appropriate potential terms in the energy of
the system. The process of DNA unzipping by applied force has been
studied in great detail
\cite{Cocco99},\cite{Cocco02},\cite{Cocco021}. The merit of these
works is that many of the theoretical conclusions, gained on the
basis of rather simple network models, are in agreement with the
experimental results.

In the current study of the nonlinear dynamics related with the
opening process and elasticity features of  DNA molecules our aim
is to examine the energy exchange processes and the relaxation
dynamics in DNA molecules after
 they have been brought into a non-equilibrium
conformation by applied tension. Such an investigation is
associated with recent mechanical experiments performed with
single  DNA molecules \cite{Clausen}-\cite{Nelson} forced away
from their equilibrium conformations. Following force applications
energy redistribution within the DNA molecule takes place such
that a (new) equilibrium conformation is attained
\cite{Bockelmann1},\cite{Clausen},\cite{Crisona},\cite{Leger}. We
focus our interest on the elasticity properties and the relaxation
dynamics of two different double-stranded DNA configurations,
namely the planar untwisted (ladder-like) system and the twisted
one arising when the two strands of the molecule are coiled around
its molecular axis. In particular we address the questions in
which way these two different  DNA molecules react to imposed
stress and how fast the equilibration process takes place for
them. Furthermore, we compare the original equilibrium structures
with those newly attained after release from the exerted
mechanical stress. Regarding the biological functioning of DNA
their elasticity properties are of importance for instance for the
creation of a denatured bubble appearing prior to strand
separation making the DNA accessible to the transcription and
replication processes. Compared to previous theoretical studies of
the unzipping dynamics of DNA induced by strong forces yielding
complete strand separation \cite{Cocco99},\cite{Cocco02} we remark
that for the current study we rather concentrate on mechanically
stressed double-stranded DNA which gets forced away from the
equilibrium configuration, however, only with beginning strand
separation. From a biophysical point of view this also resembles
the situation when the DNA molecule is inflicted to internal
stress changes occurring in the living cell which alone is not strong
enough to cause advanced bond breaking.
Furthermore, the interaction between
DNA and proteins
such as enzymes   can exert forces that initiate the local
opening of the DNA molecule.

The paper is organized as
follows: In the second section we describe our network model for
the structure of double-stranded DNA. The third section deals with
the relaxation dynamics within DNA molecules forced into locally
distorted configurations. We study the similarities and
differences between the relaxation dynamics for a planar
ladder-like DNA molecule and a twisted one. Finally, we summarize
our results.

\section{Network model for the double-stranded DNA}

Our DNA model takes  the basic geometrical features of the
double-stranded form of DNA, as a polymeric molecule composed of
two strands of nucleotides, into account which are essential to
model structural deformations generated by mechanical stress
imposition and vibrational relaxation processes. Likewise the
approach in \cite{Cocco99},\cite{BCP}, we treat the
double-stranded DNA as a network of coupled oscillators
incorporating essential microscopic degrees of freedom of DNA and
the inherent interactions between them. Such a network model
approach is appealing because it can successfully describe the
mechanical behavior on the microscopic scale relevant for
biomolecular processes, but bears enough simplicity to be
tractable with not too much computational efforts. The
constituents of the oscillator network model represent the
nucleotides which are regarded as single nondeformable entities.
Thus, no inner dynamical degrees of freedom of the nucleotides are
taken into account which is justified by the time scale separation
between the small-amplitude and fast vibrational motions of the
individual atoms and the slower and relatively large-amplitude
motions of the atom groups constituting the nucleotides
\cite{Stryer}. Concerning the structural components, each
nucleotide is composed of a sugar, a phosphate and a base. The
sugar-phosphate groups of neighboring nucleotides on the same
strand are linked via covalent bonds establishing the rigid
backbone to the strand. There is a base attached to every sugar.
Since, for simplicity, we do not distinguish between the four
different types of bases, the nucleotides are considered as
identical objects of fixed mass. Two bases on opposite strands are
linked via hydrogen bonds holding the two strands of DNA together.

\begin{figure}
  \begin{center}
    \includegraphics[width=\singlefig]{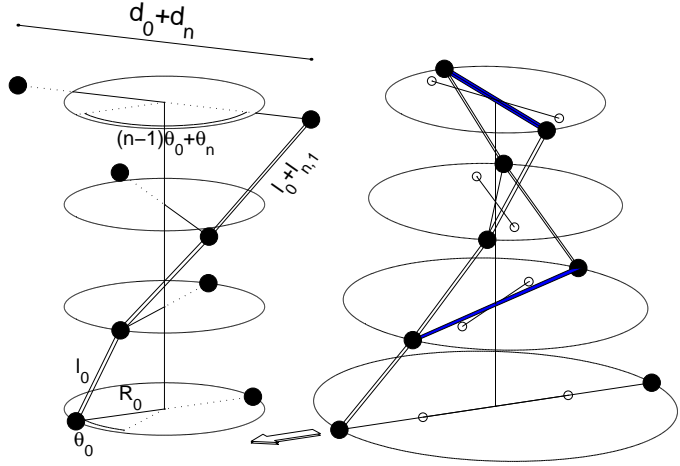}
  \caption{Left: Geometry of the twisted double-stranded
 DNA model showing the variables. Right: The DNA molecule at the end of the
pulling process with exaggerated distortions.} \label{fig:fig1}
\end{center}
\end{figure}

 For computational convenience we cast the double helix
structure in a Cartesian coordinate system whose $z-$axis
coincides  with the central helix axis as sketched in Fig.\,$1$.
The base pairs are situated in planes perpendicular to the central
helix axis and the vertical distance between two consecutive
planes is given by $h$. For the equilibrium configuration each
base has equilibrium coordinates $x_{n,i}^{(0)}$, $y_{n,i}^{(0)}$
and $z_{n,i}^{(0)}$. The index pair $(n,i)$ labels the $n$-th base
on the $i-$th strand with $i=1,2$ and  $1\leqq n\leqq N$\,, where
$N$ is the number of base pairs considered.
 The equilibrium distance between two
bases within a base pair, $d_0$, is determined by
\begin{equation}
d_0=\sqrt{\left(\,d_n^x\,\right)^2\,+\,
\left(\,d_n^y\,\right)^2}\,,
\end{equation}
where  $d_n^x=\,x_{n,1}^{(0)}-x_{n,2}^{(0)}$ and
$d_n^y=\,y_{n,1}^{(0)}-y_{n,2}^{(0)}$ are the projections of the
line connecting the two bases on the $x,y-$axes of the coordinate
system.
The deviations $d_n$ from $d_0$ through displacements, $x_{n,i}$,
$y_{n,i}$ and $z_{n,i}$, of the bases from their equilibrium
positions, $x_{n,i}^{(0)}$, $y_{n,i}^{(0)}$ and $z_{n,i}^{(0)}$,
are expressed as
\begin{equation}
d_n=\sqrt{\left(\,d_n^x+x_{n,1}-x_{n,2}\right)^2\,+\,
\left(\,d_n^y+y_{n,1}-y_{n,2}\right)^2
\,+\,\left(\,z_{n,1}-z_{n,2}\,\right)^2}-d_0\,.
\end{equation}
The relative twist between two consecutive base pairs, being
rotated around the central axis, is given by an angle
$\theta_{0}$. For later use we introduce the quantity
\begin{equation}
\theta_n=\arctan\,\frac{d_n^y+y_{n,1}-y_{n,2}}{d_n^x+x_{n,1}-x_{n,2}}
+2m\pi\,,
\end{equation}
as the angle between the $x-$axis (as the reference direction) and
the line connecting two (displaced) bases of a base pair measuring
the alignment of the associated H-bridge. $m$ is an integer to
assure monotonicity of $\theta_n$ with respect to $n$.

The three-dimensional equilibrium distance between two adjacent
bases on the same strand is given by
\begin{equation}
l_0=\sqrt{\left(\,x_{n,i}^{(0)}-x_{n-1,i}^{(0)}\right)^2\,+\,
\left(\,y_{n,i}^{(0)}-y_{n-1,i}^{(0)}\right)^2+h^2}\,,
\end{equation}
and deviations from $l_0$ are determined by
\begin{eqnarray}
l_{n,i}&=&\left\{\left(\,L_{n,i}^x+x_{n,i}-x_{n-1,i}\right)^2\,+\,
\left(\,L_{n,i}^y+y_{n,i}-y_{n-1,i}\right)^2\right.\nonumber\\
&+&\left.\,\left(\,h+z_{n,i}-z_{n-1,i}\,\right)^2\right\}^{1/2}-l_0\,,
\end{eqnarray}
with $L_{n,i}^x=x_{n,i}^{(0)}-x_{n-1,i}^{(0)}$ and
$L_{n,i}^y=y_{n,i}^{(0)}-y_{n-1,i}^{(0)}$.

 Our model  Hamiltonian is then given by
\begin{equation}
H=E_{kin}+V_{h}+V_{c}+V_{s}+V_{l}\,.\label{eq:Hdna}
\end{equation}
$V_{h}$ and $V_{c}$ represent the potential energy part for the
H-bonds and the covalent bonds, respectively. The deformations of
the hydrogen bonds corresponding to
 deviations  $d_n(t)$ from the equilibrium bond length
$d_0$ are described by vibrations in a Morse potential which is
given by
\begin{equation}
U_{h}= D\,\sum_{i=1,2}\,\sum_{n=1}^{N}\,
\left[\,\exp\left(-\alpha\, d_n\right)\,-1\,\right]^2\,.
\label{eq:Uhyd}
\end{equation}
D is the depth of  the Morse potential and $\alpha$ is the range
parameter. For the sake of simplicity we  do not distinguish
between the two different pairings in DNA, namely the G-C and the
A-T pairs. The former pair involves three hydrogen bonds while the
latter involves only two. Thus our model applies to pure
poly(dG-dC) and poly(dA-dT) DNA sequences.

The covalent bonds
between the sugar-phosphate groups of neighboring nucleotides on a
strand are rather strong and rigid (with bond energies of the
order of $2-10\,eV$) in comparison with the weak and flexible hydrogen bonds (with bond
energies of the order of $0.04-0.3\,eV$).
Therefore, it seems justified to treat the
potential of the covalent bonds, simulated by elastic rods, in the
harmonic approximation given by
\begin{equation}
U_{c}=K\,\sum_{i=1,2}\,\sum_{n=1}^{N}\,l_{n,i}^2\,,
\label{eq:Uback}
\end{equation}
where $K$ is the elasticity coefficient.

The potential term $V_{s}$ takes stacking effects into account
which impede that one base slides over another \cite{Stryer}. For
the form of $V_{s}$ we adopt the one used in  \cite{Cocco99}
\begin{equation}
V_{s}=\frac{S}{2}\,\sum_{n,i}\,(d_{n,i}-d_{n-1,i})^2\,
\exp[-\beta(d_{n,i}+d_{n-1,i})]\,.
\end{equation}
The decrease of the molecular packing with substantially increased
base pair opening is accounted for by the exponential factor with
attenuation parameter $\beta$. As the changes of the axial
distance between adjacent base planes is concerned we assume that
the supposedly small longitudinal helix deformations can be
modeled by a harmonic elasticity potential term given by
\begin{equation}
V_{l}=\frac{C}{2}\,\sum_{n,i}\,(z_{n,i}-z_{n-1,i})^2\,.
\end{equation}
The kinetic energy is determined by
\begin{equation}
E_{kin}=\frac{1}{2m}\,\sum_{i=1,2}\,
\sum_{n=1}^{N}\,\left[\,\left(p_{n,i}^{(x)}\right)^2
+\left(p_{n,i}^{(y)}\right)^2
+\left(p_{n,i}^{(z)}\right)^2\right]\,,\label{eq:Ekin}
\end{equation}
where $m$ is the mass of a base and $p_{n,i}^{(x,y,z)}$ denote the
$(x,y,z)-$component of its momentum.

In order to assign values to the various parameters we note that
the geometrical parameters of the equilibrium configuration are
well known \cite{Stryer}. The rotation angle for the twisted
configuration is $\theta_0=36^\circ$, the distance between base
pair planes is $h=3.4$\,\AA\,, and the inter-base distance is
$d_0=20$\,\AA. (For the twisted configuration  $d_0$ corresponds
to the diameter of the helix.) The ladder-like system is
completely untwisted, viz. $\theta_0=0$. For the average mass of
one nucleotide we use $M=4.982\times10^{-25}kg$. Like Barbi {\it
et al} \cite{BCP} we set $\alpha=4.45$\,\AA$^{-1}$, $D=0.04\,eV$,
and $K=1.0\,eV$\,\AA$^{-2}$. Concerning the parameters $C$ and $S$
there is little experimental evidence from which an estimate of
them could be inferred. In \cite{Barbithesis} the value of the
parameter $S$ has been set to $S=2K$ which we adopt here.
(Alternatively, in a model approach of DNA denaturation dynamics
\cite{Cocco99} the parameters $S$ and $C$ were specified to fit
the melting temperature of certain DNA polymers.) The parameter
$C$ is treated as adjustable.  Furthermore we use for the value of
the attenuation parameter $\beta=0.5$\,\AA$^{-1}$ \cite{Cocco99}.

With a suitable time scaling  $t\rightarrow \sqrt{D
\alpha^2/m}\,t$
 and dividing the original Hamiltonian by $D$
 one passes to a dimensionless formulation with quantities:
\begin{eqnarray}
 \tilde{x}_{n,i}&=&\alpha
x_{n,i}\,, \,\,\,\, \tilde{y}_{n,i}=\alpha y_{n,i}\,,\,\,\,\,
\tilde{z}_{n,i}=\alpha z_{n,i}
\\
\tilde{p}_{n,i}^{(x)}&=&\frac{p_{n,i}^{(x)}}{\sqrt{mD}}\,,\,\,\,\,
\tilde{p}_{n,i}^{(y)}=\frac{p_{n,i}^{(y)}}{\sqrt{mD}} \,,\,\,\,\,
\tilde{p}_{n,i}^{(z)}=\frac{p_{n,i}^{(z)}}{\sqrt{mD}} \,,\\
\tilde{C}&=&\frac{C}{\alpha^2 D}\,,\,\,\,\,
\tilde{K}=\frac{K}{\alpha^2 D}\,,\,\,\,\,
\tilde{S}=\frac{S}{\alpha^2 D}\,,\\ \tilde{d}_{n}&=&\alpha\,
d_{n}\,,\,\,\,\,
\tilde{r}_{0}=\alpha\,r_{0}\,,\,\,\,\,\tilde{h}=\alpha\,h
\,,\,\,\,\,\tilde{\beta}=\frac{\beta}{\alpha} \,.
\end{eqnarray}
In the following, we omit the tildes.

The equations of motion are derived from the Hamiltonian presented
in Eqs. (\ref{eq:Hdna})-(\ref{eq:Ekin}) and read as
\begin{eqnarray}
\dot{x}_{n,i}&=&p_{n,i}^{(x)}\,,\label{eq:xdot}\\
\dot{p}_{n,i}^{(x)}&=&2\left[\,\exp(-d_n)-1\,\right]\,\exp(-d_n)\,
\frac{\partial d_n}{\partial x_{n,i}}\nonumber\\
&-&2K\left[\,l_{n,i}\frac{\partial l_{n,i}}{\partial x_{n,i}}+
l_{n+1,i}\frac{\partial l_{n+1,i}}{\partial
x_{n,i}}\,\right]\nonumber\\
&-&S\,\left\{\left[\,1-\frac{\beta}{2}(d_{n}-d_{n-1})\,\right]\,
(d_{n}-d_{n-1})\,\exp[-\beta\,(d_{n}+d_{n-1})]\right.\nonumber\\
&-&\left.\left[\,1+\frac{\beta}{2}(d_{n+1}-d_{n})\,\right]\,
(d_{n+1}-d_{n})\,\exp[-\beta\,(d_{n+1}+d_{n})]\right\}\nonumber\\
&\times&\,\frac{\partial d_n}{\partial x_{n,i}}\,,\\
\dot{y}_{n,i}&=&p_{n,i}^{(y)}\,,\label{eq:ydot}\\
\dot{p}_{n,i}^{(y)}&=&2\left[\,\exp(-d_n)-1\,\right]\,\exp(-d_n)\,
\frac{\partial d_n}{\partial y_{n,i}}\nonumber\\
&-&2K\left[\,l_{n,i}\frac{\partial l_{n,i}}{\partial y_{n,i}}+
l_{n+1,i}\frac{\partial l_{n+1,i}}{\partial
y_{n,i}}\,\right]\nonumber\\
&-&S\,\left\{\left[\,1-\frac{\beta}{2}(d_{n}-d_{n-1})\,\right]\,
(d_{n}-d_{n-1})\,\exp[-\beta\,(d_{n}+d_{n-1})]\right.\nonumber\\
&-&\left.\left[\,1+\frac{\beta}{2}(d_{n+1}-d_{n})\,\right]\,
(d_{n+1}-d_{n})\,\exp[-\beta\,(d_{n+1}+d_{n})]\right\}\nonumber\\
&\times&\,\frac{\partial d_n}{\partial
y_{n,i}}\,,\label{eq:pydot}\\
\dot{z}_{n,i}&=&p_{n,i}^{(z)}\,,\label{eq:zdot}\\
\dot{p}_{n,i}^{(z)}&=&2\left[\,\exp(-d_n)-1\,\right]\,\exp(-d_n)\,
\frac{\partial d_n}{\partial z_{n,i}}\nonumber\\
&-&2K\left[\,l_{n,i}\frac{\partial l_{n,i}}{\partial z_{n,i}}+
l_{n+1,i}\frac{\partial l_{n+1,i}}{\partial
z_{n,i}}\,\right]\nonumber\\
&-&S\,\left\{\left[\,1-\frac{\beta}{2}(d_{n}-d_{n-1})\,\right]\,
(d_{n}-d_{n-1})\,\exp[-\beta\,(d_{n}+d_{n-1})]\right.\nonumber\\
&-&\left.\left[\,1+\frac{\beta}{2}(d_{n+1}-d_{n})\,\right]\,
(d_{n+1}-d_{n})\,\exp[-\beta\,(d_{n+1}+d_{n})]\right\}\nonumber\\
&\times&\,\frac{\partial d_n}{\partial z_{n,i}}\nonumber\\
&-&C\left(\,2\,z_{n,i}-z_{n+1,i}-z_{n-1,i}\,\right)\,,\label{pzdot}
\end{eqnarray}
with the derivatives
\begin{eqnarray}
\frac{\partial d_n}{\partial
x_{n,i}}&=&\frac{(-1)^{i+1}\,(d_n^x+x_{n,1}-x_{n,2})}
{d_n+d_0}\,,\\ \frac{\partial l_{n,i}}{\partial x_{n,i}}&=&
\frac{L_x+x_{n,i}-x_{n-1,i}}{l_{n,i}+l_0}\,,
\end{eqnarray}
and the equivalent expressions for ${\partial d_n}/{\partial
y_{n,i}}$, ${\partial l_{n,i}}/{\partial y_{n,i}}$, ${\partial
d_n}/{\partial z_{n,i}}$ and ${\partial l_{n,i}}/{\partial
z_{n,i}}$.

The values of the scaled parameters are given by $K=0.683$,
$r_0=44.50$, $h=15.13$ and $l_0=31.39$. One time unit of the
scaled time corresponds to $0.198\,ps$ of the physical time.

\section{Relaxation dynamics of mechanically stressed DNA}

\subsection{Pulling process in the ladder and twisted
configuration}

 We study the energy exchange process taking place
after the double-stranded DNA has been forced mechanically into a
non-equilibrium configuration. In more detail, we consider the
situation when at one end (referred to as the {\it pulled end}) of
short DNA molecules consisting of $10$ up to $30$ base pairs the
two strands get pulled laterally in opposite directions by an
external force (exerted e.g. when the two strands of one end of
the molecule are attached to two different solid supports that are
progressively separated \cite{Bockelmann}) so that they are
dragged apart leading to gradual opening of the double-stranded
chain.
The pulled end is formed by the base pairs with index $n=1$ and
the rest of the molecule with $n>1$ is located above it in the $z$
direction. Therefore a positive value of
$z_{n,i}-z_{n-1,i}$  will
represent a local $z$-stretching of the strand $i$.
In order to mimic the pulling process we assume that the pulling
force is directed solely along the $x-$direction and  the
$x-$coordinates of the terminal bases are shifted uniformly with a
pulling velocity of $v_p=20\mu m/s$
 while the corresponding $y$
 and $z$ coordinates are kept fixed at
 $y_{1,i}=z_{1,i}=0$.

Moreover, due to the constraint imposed by the backbone rigidity
there result deformations not only at the pulled end but also for
the rest of the helix. In general, the extent of lateral
deformations subsides along the strands from the pulled towards
the free end. The motion of the pulling unit is assumed to
terminate when the lateral elongation at the pulled end has
reached a certain value, taken as
 $d= 2.31\,\AA$,  so that the
corresponding terminal hydrogen bond is broken. Moreover, the
next-neighboring hydrogen bond is broken too and the third one is
already considerably stretched. The typical resulting deformation
pattern of the double stranded DNA is depicted in Figs.\,$2$ and
$3$ for the ladder and twisted configuration, respectively. (In
the Figures energy is expressed in terms of $eV$, the unit for
length is $\AA$ and the time is given in $ns$.)

Comparing the resulting DNA deformations we note that  for both,
the ladder and the twisted configuration, virtually equal lateral
elongations are obtained. However, unlike the ladder configuration
the twisted one experiences additional changes of the twist
angles. There arises an overall decrease in the twist angle and
the further a base pair is apart from the pulled end the more is
the corresponding twist angle reduced (cf. Fig.\,$3\,(b)$) leading
to an untwist of the helix. Nevertheless, with growing distance
from the pulled end the difference between two consecutive angle
changes becomes smaller. Thus the relative twist angle between two
adjacent base planes diminishes towards the free end.

\begin{figure}
\begin{tabular}{cc}
    \includegraphics[angle=-90,width=0.5\textwidth]{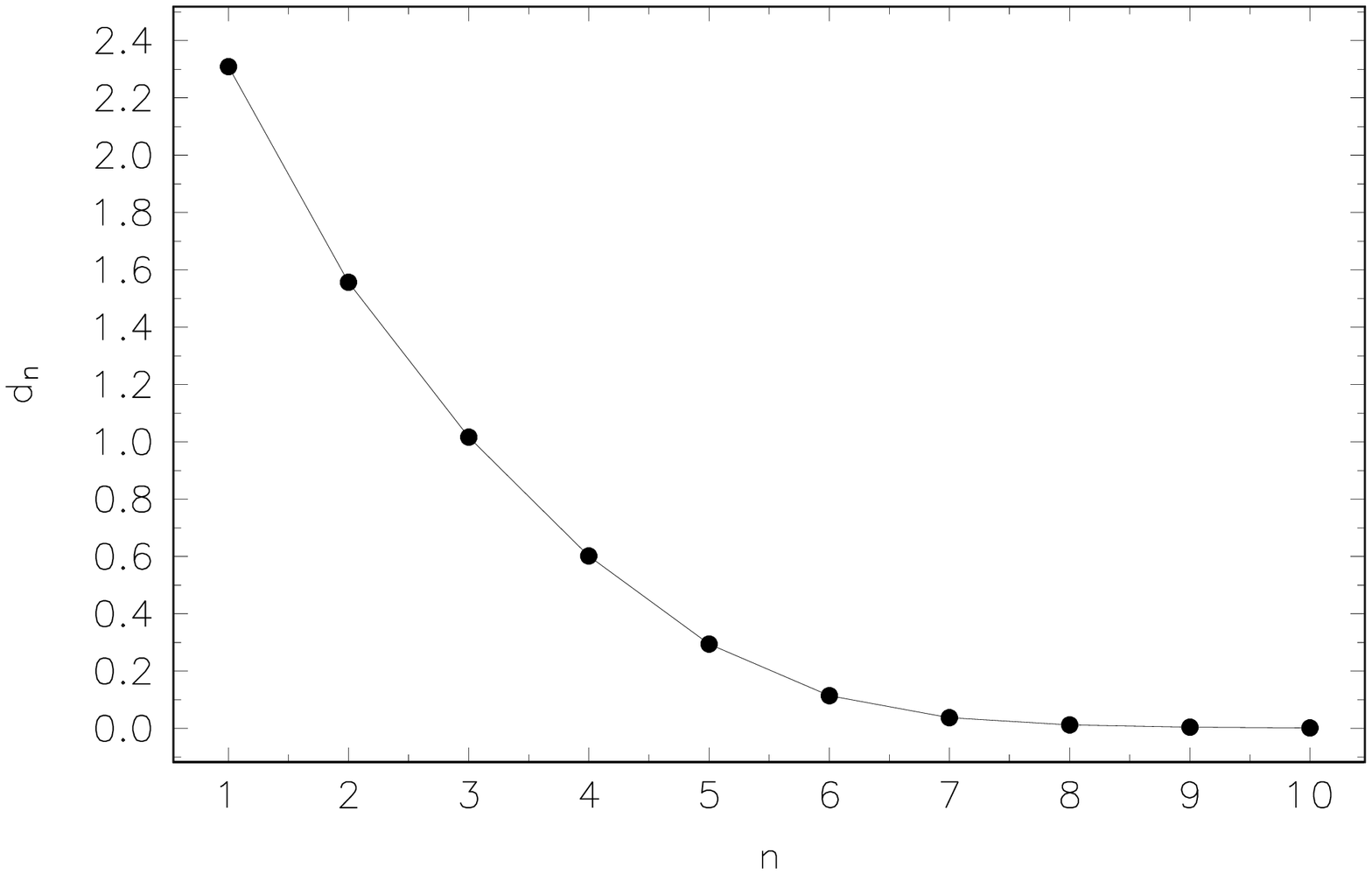}&
    \includegraphics[angle=-90,width=0.5\textwidth]{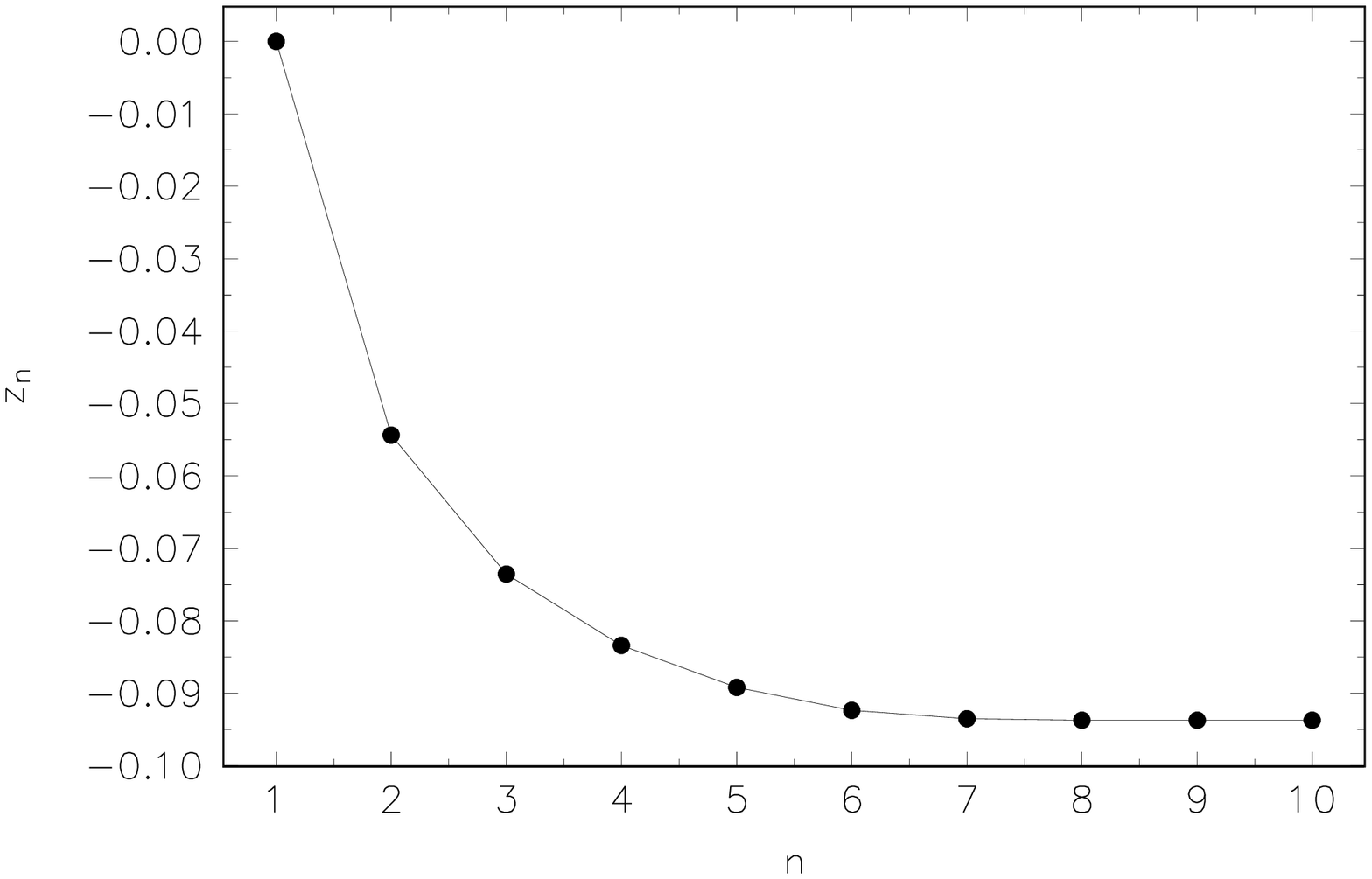}
\end{tabular}
  \caption{Deformation pattern at the end of the
pulling process for the ladder system. (a) The lateral
deformations $d_n$. (b) The deformations in axial direction $z_n$.
} \label{fig:fig2}
\end{figure}

\begin{figure}
  \begin{center}
\begin{tabular}{cc}
    \includegraphics[angle=-90,width=0.5\textwidth]{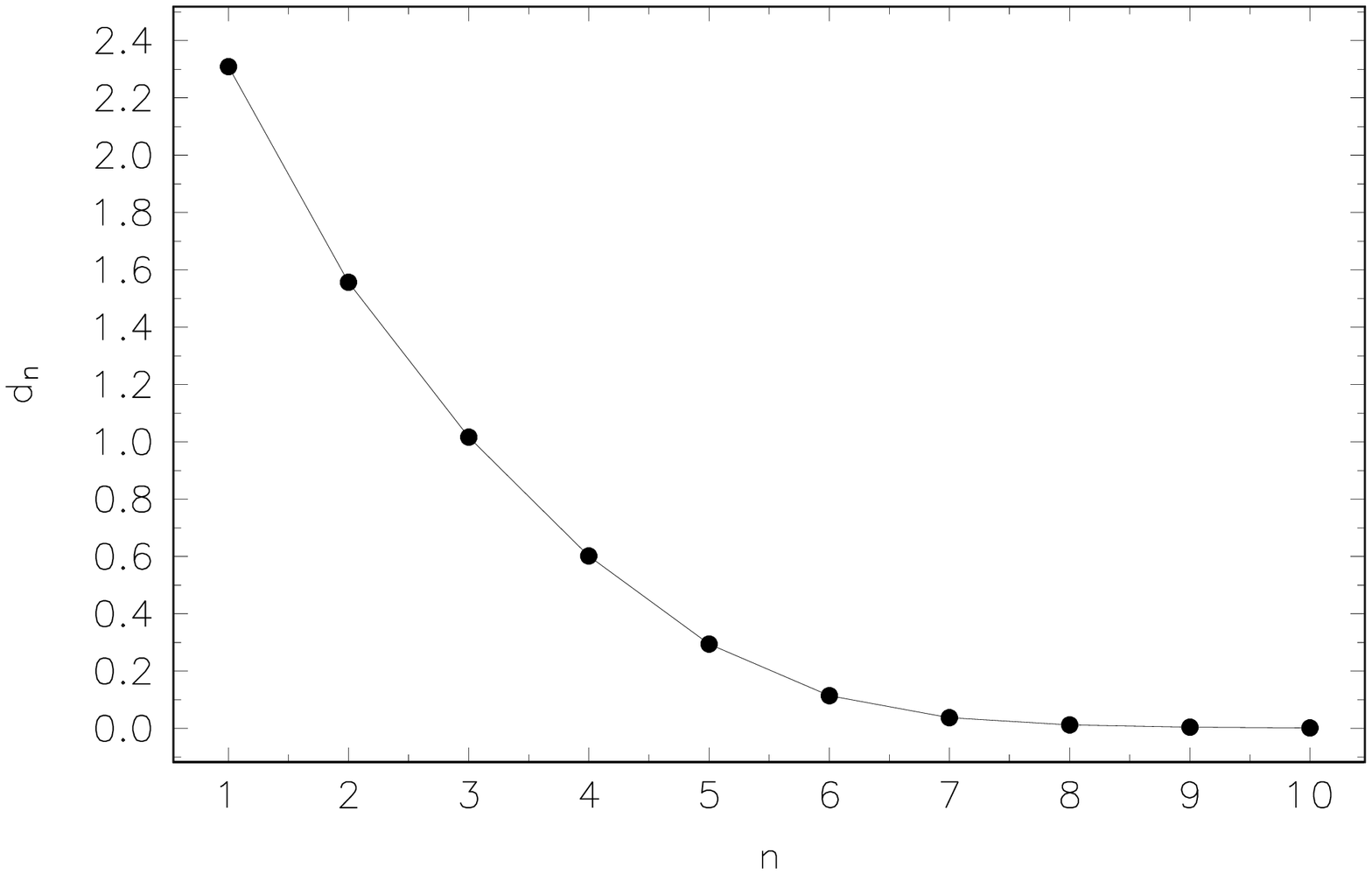}&
   \includegraphics[angle=-90,width=0.5\textwidth]{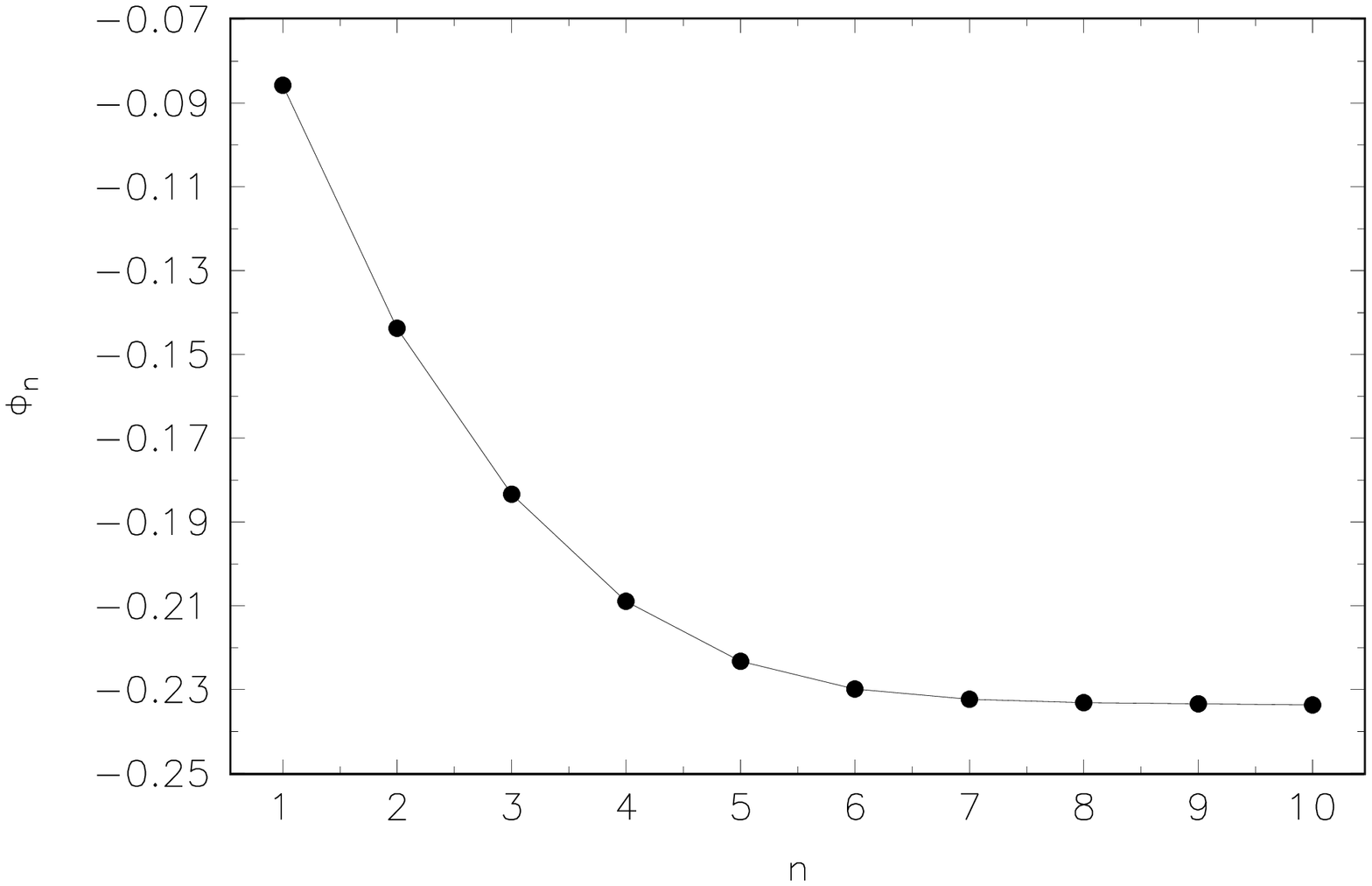}
\end{tabular}
   \includegraphics[angle=-90,width=0.5\textwidth]{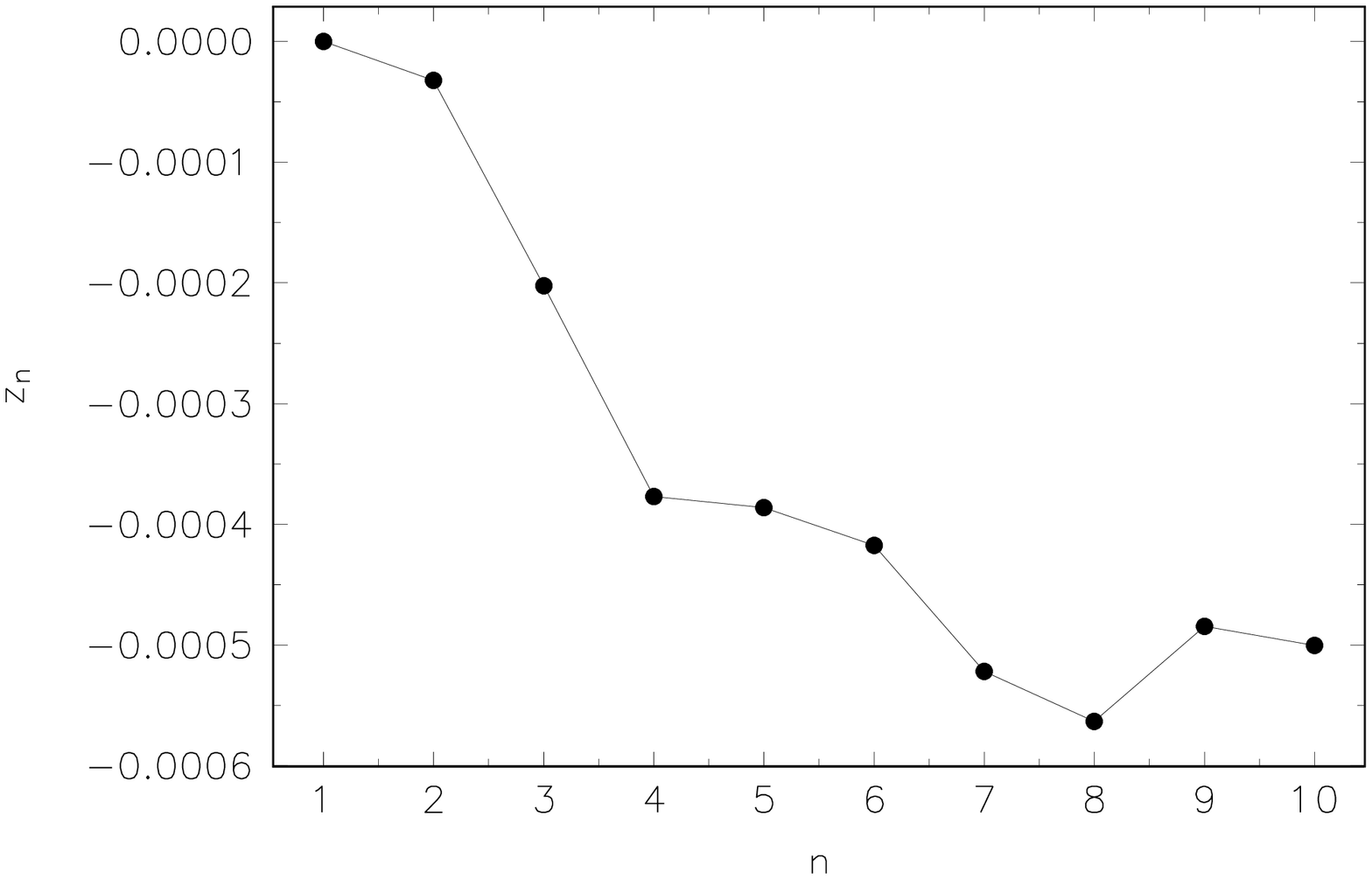}
  \caption{Deformation pattern at the end of the
pulling process for the twisted system. (a) The lateral
deformations $d_n$.(b) The twist angle changes,
$\phi_n=\theta_n-(n-1)\theta_0$.(c) The deformations in axial
direction $z_n$.} \label{fig:fig3}
\end{center}
\end{figure}

While for the twisted system the axial deformations are negligible
  (see Fig.\,$3\,(c)$)
  we note that for the ladder configuration
there results noticeable diminution of the axial equilibrium
distances between neighboring base planes which is strongest at
the free end and weakens gradually towards the pulled end
 (see Fig.\,$2\,(b)$).
The relations between the partial deformation energies
 for the twisted system
are shown in Fig.\,$4$. The major contributions stem from the
Morse potential energy associated with the H-bridge deformations
and the stacking interaction energy both exceeding the energetic
content of the covalent bond and axial distance change energy,
respectively by four orders of magnitude. Typical for DNA
molecules is that there exist different elasticity regimes, that
is the degree of the deformation depends not overall linearly on
the force (see e.g. \cite{Clausen}) which is reflected here in the
nonlinear character of the temporal behavior of the deformation
energies.

\begin{figure}
  \begin{center}
    \includegraphics[angle=-90,width=\singlefig]{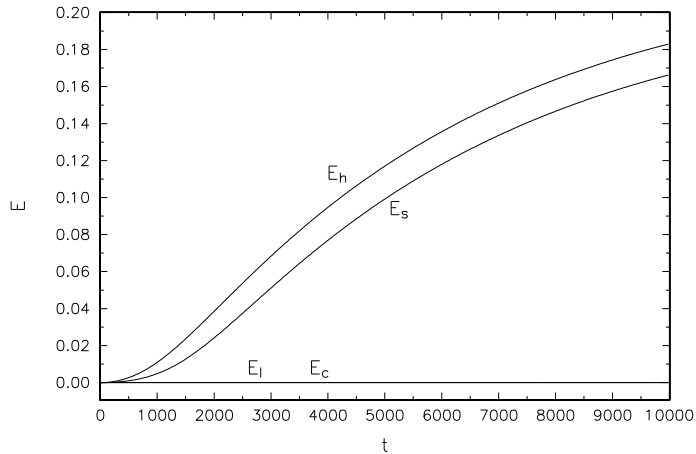}
  \caption{Pulling process for the twisted system: The temporal behavior
of the partial deformation energies as indicated in the figure. }
\label{fig:fig4}
\end{center}
\end{figure}

\subsection{Relaxation processes}
After the DNA molecule has been brought into a {\it
non-equilibrium} configuration with imposed mechanical tension we
explore now the vibrational dynamics of the helix cage after the
pulled end has been released from the forcing lever. Hence, there
is no external  mechanical constraint enforced on the double helix
any longer. For the then isolated double helix we expect that
energy redistribution processes between the various vibrational
degrees of freedom take place targeting for the attainment of a
 new stationary configuration.

We discuss the ladder system and the twisted one separately and
start with the former case for stacking interaction strength
$S=1.26$.

We integrated the set of coupled equations
(\ref{eq:xdot})-(\ref{eq:pydot})
  (with free boundary conditions)
with a fourth-order Runge-Kutta method while the accuracy of the
computation was checked through the conservation of the total
energy.

To gain insight into the energy redistribution process we monitor
the temporal behavior of the potential energies associated with
the deformations of the hydrogen and the covalent bonds,
respectively, and changes of the axial elasticity and stacking
interaction energy, respectively.
 Fig.\,$5$.
 illustrates how
the total deformation energy of $0.35\,eV$, which has been
deposited in the DNA molecule, is shared between the various
degrees of freedom in the course of time.
 Generally, we observe that in an initial interval of
 $\sim7000\,ns$, hereafter referred to as the {\it redistribution
time}, pronounced migration of energy takes place. Regarding the
Morse potential energy we note that during this initial interval
the amplitude oscillates wildly in a range reaching from
$0.05\,eV$ up to $0.30\,eV$. Afterwards the potential energy
performs lower-amplitude oscillations around the mean value
$0.1\,eV$ indicating that a quasi-equilibrium regime is attained.
Apparently, the amount of energy by which  the Morse potential
energy gets reduced (almost $50\,\%$ of its initial value) has
been redistributed into other forms of the deformation energy.
Likewise the Morse potential energy, the stacking interaction
energy experiences also losses which manifests that the stretched
H-bonds relax and come close to their equilibrium lengths.
Subsequent to a drop, the stacking interaction energy performs
oscillations of fairly large amplitudes throughout the time around
the mean value of $0.11\,eV$. Hence, the reduced deformation
energy of the H-bonds flows into
 the covalent bond energy and the axial distance change
energy exhibit analogous temporal behavior, i.e. both rise on the
average during the redistribution time and approach their
quasi-equilibrium values.  Eventually, the energies $E_c$ and
$E_l$ oscillate around their respective mean value. In other
words, the H-bond deformation energy is reduced and its loss is
deposited into such forms of the energy which contribute mainly to
enhanced deformations of the covalent bonds. However, the energy
associated with the axial distance changes remains still very
small compared to all other deformation energies. We remark that
the mean value of the covalent bond energy has grown up to
$0.06\,eV$ and the peaks of the oscillations are even in the range
of the Morse potential and the stacking interaction energy giving
evidence for major distortions of the covalent bonds.

\begin{figure}
\begin{tabular}{cc}
    \includegraphics[angle=-90,width=0.5\textwidth]{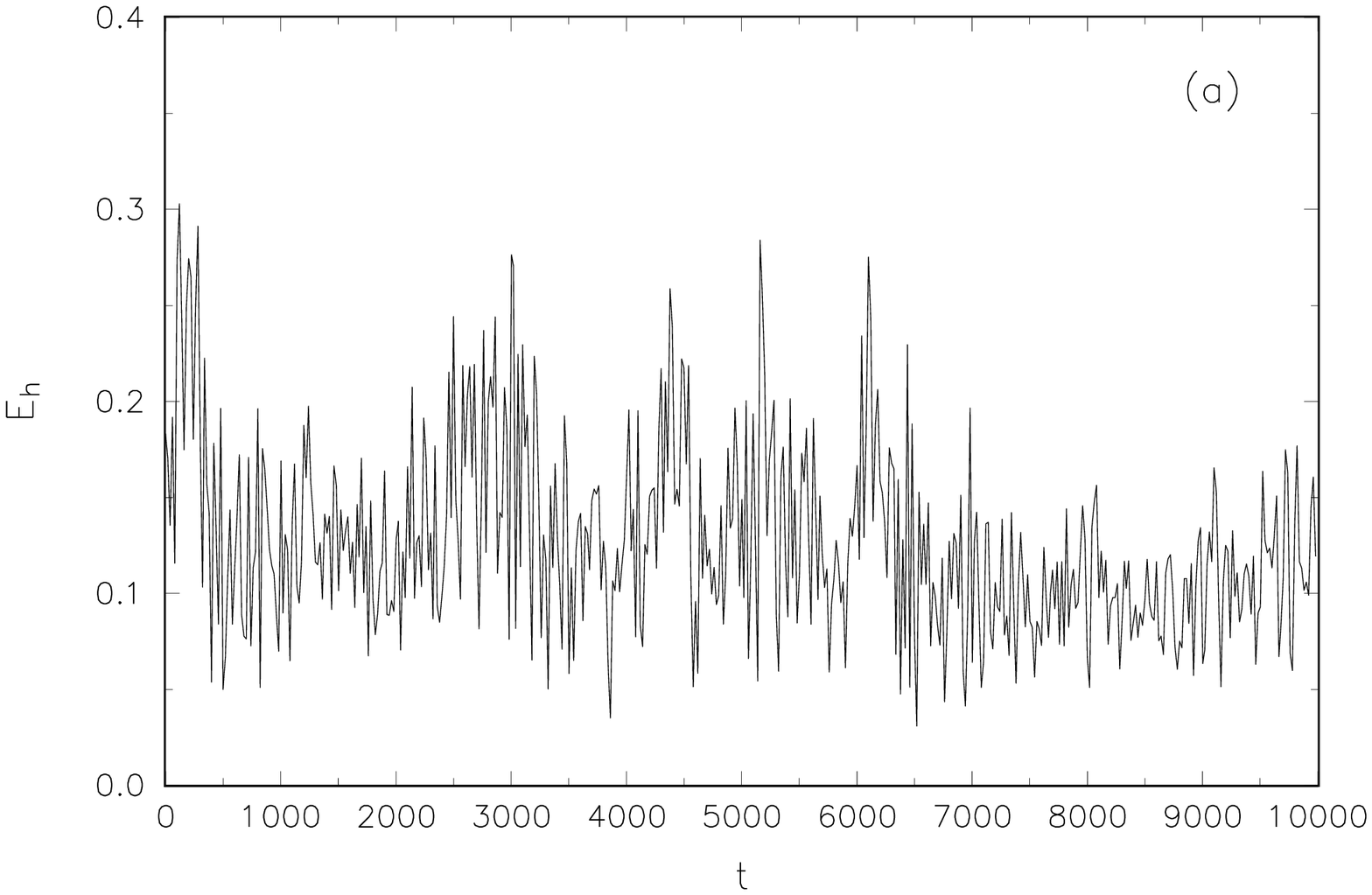}&
    \includegraphics[angle=-90,width=0.5\textwidth]{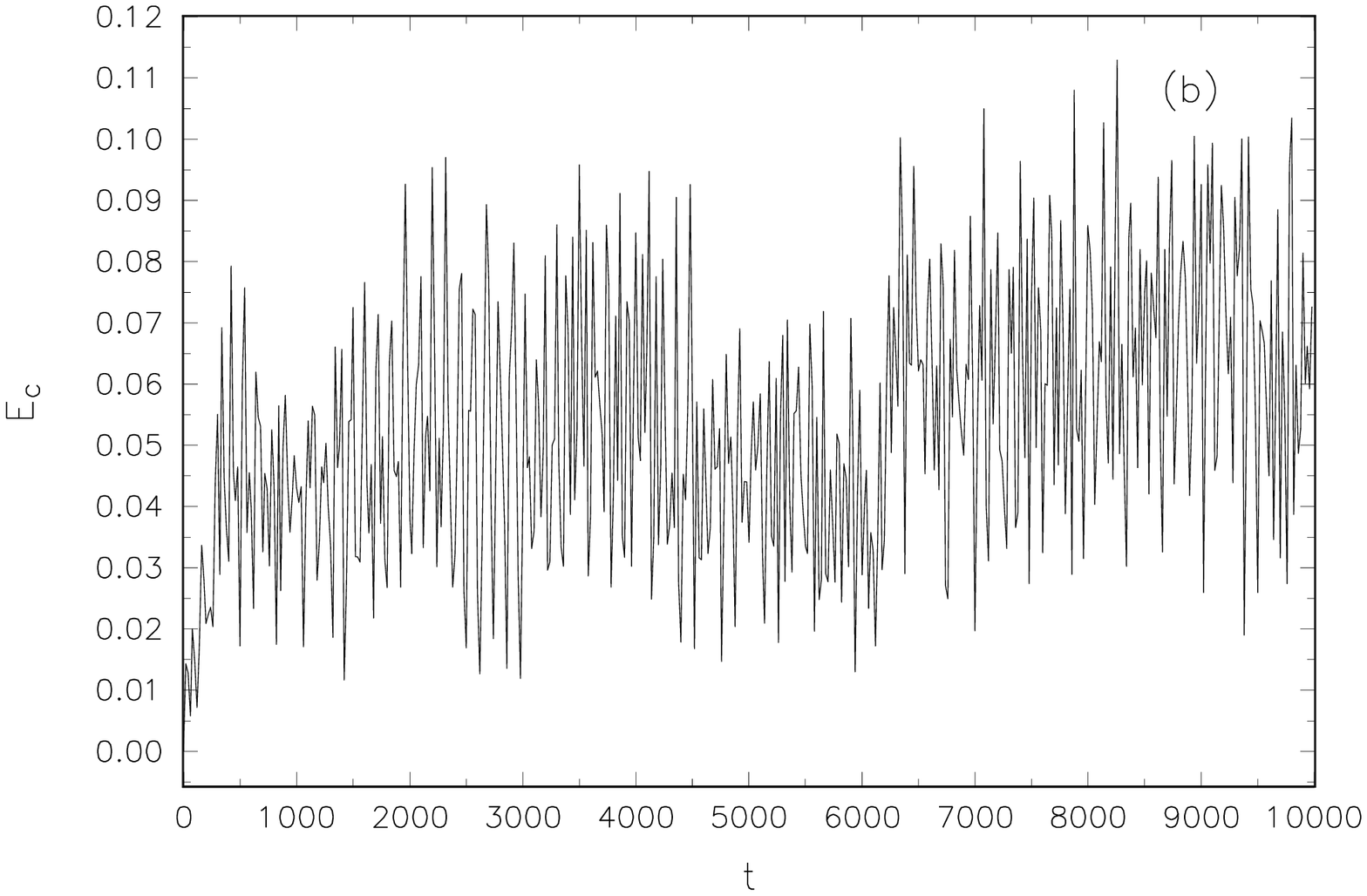}\\
    \includegraphics[angle=-90,width=0.5\textwidth]{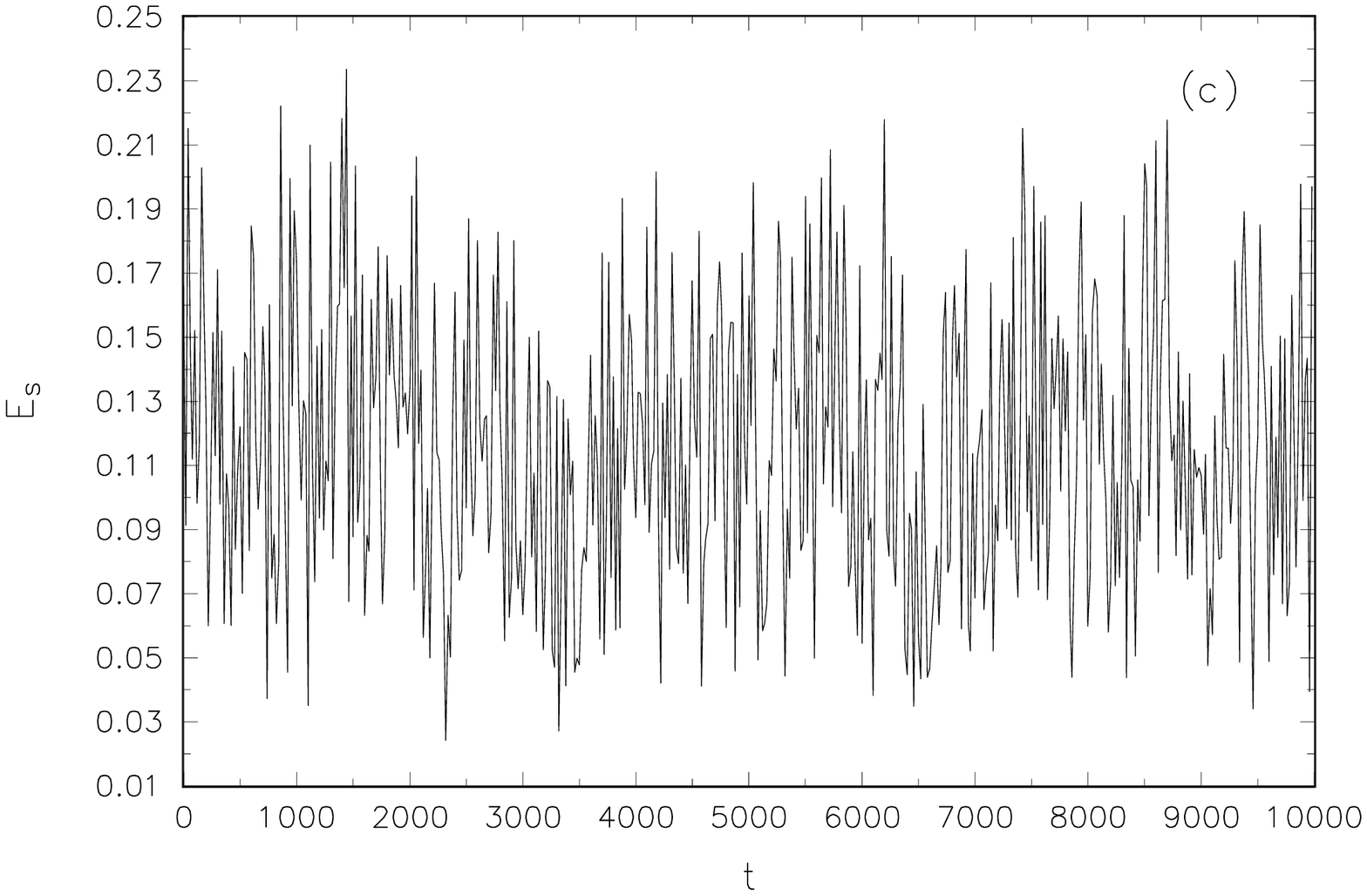}&
    \includegraphics[angle=-90,width=0.5\textwidth]{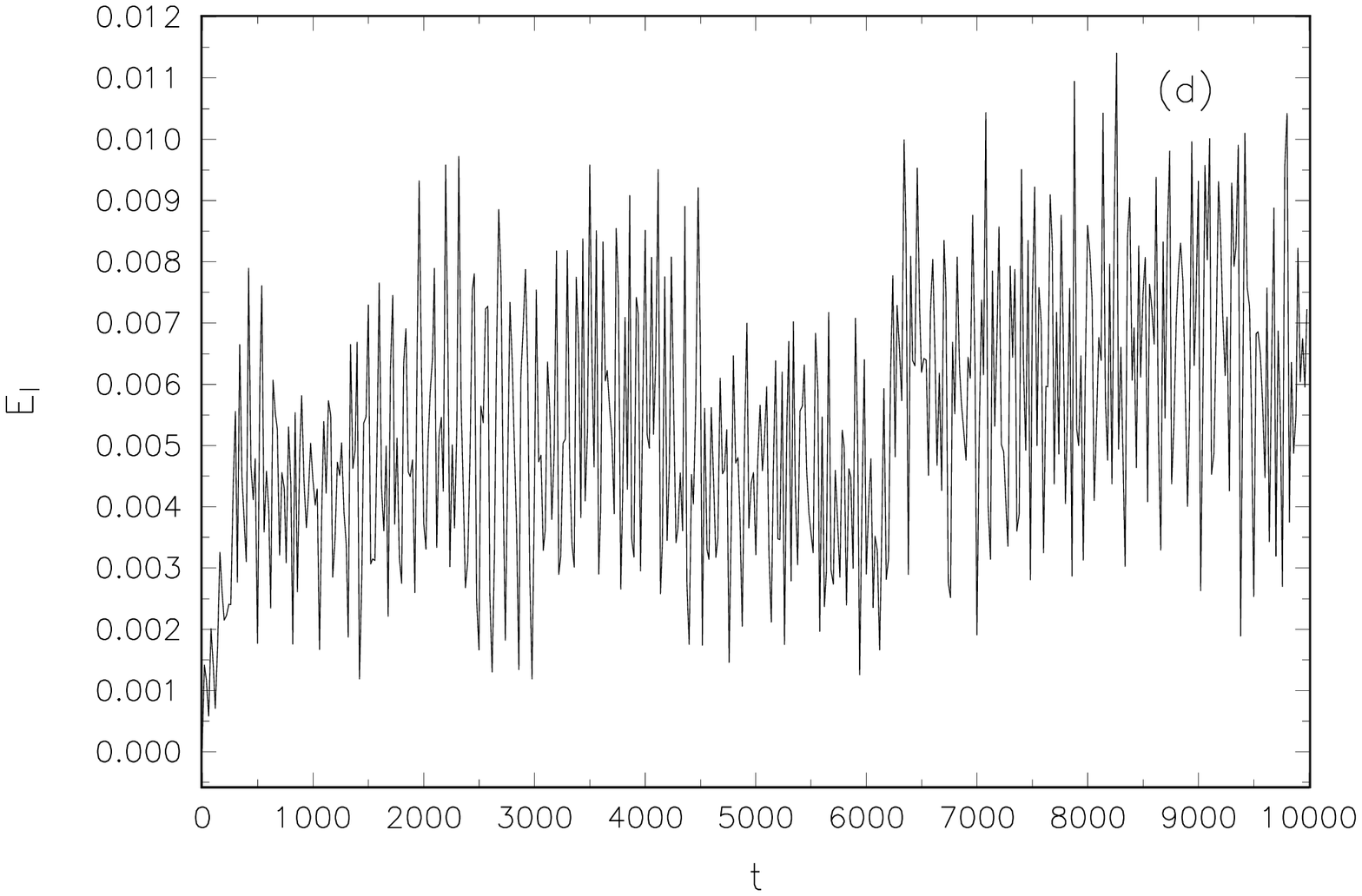}
    \end{tabular}
  \caption{Relaxation dynamics for the ladder
system: Time evolution of the partial potential energies: (a)
$V_h$, (b) $V_c$. (c) $V_s$, and (d) $V_l$.
 } \label{fig:fig5}
\end{figure}

The energy redistribution processes in the twisted case are
illustrated in Fig.\,$6$. Compared to the  behavior in the
preceding ladder case this time the  Morse potential energy
descends already within  $2000\,ns$ down to its mean value pointing to a
faster equilibration process. Nevertheless, there appear
occasional bursts of the amplitude. The stacking interaction
energy  has rapidly dropped down from the initial value $\sim
0.17\,eV$ to $\sim 0.02\,eV$ after only $\sim 30\,ns$.
Subsequently, the stacking interaction energy exhibits
oscillations around a mean value of $0.55\,eV$. However, as the
covalent bond energy is concerned we find that it increases
substantially and the eventually attained mean value exceeds even
that of the Morse potential energy. This behavior has to be
distinguished from the redistribution process in the ladder case
where, in the quasi-equilibrium regime,  the deformation energy is
primarily stored as Morse potential and stacking interaction
energy implying that H-bridge deformations play a larger role than
any other DNA deformations. Moreover, for the twisted case we
observe also a much more drastic increase in the axial distance
change energy and the covalent bond energy relative to the ladder
case which is associated with intensified deformations of the
helix backbone. Hence, the process of energy redistribution
connected with the approach of a (quasi) equilibrium state
proceeds differently in the ladder case and the twisted one
causing also in the outcome distinct structural modifications.

\begin{figure}
\begin{tabular}{cc}
    \includegraphics[angle=-90,width=0.5\textwidth]{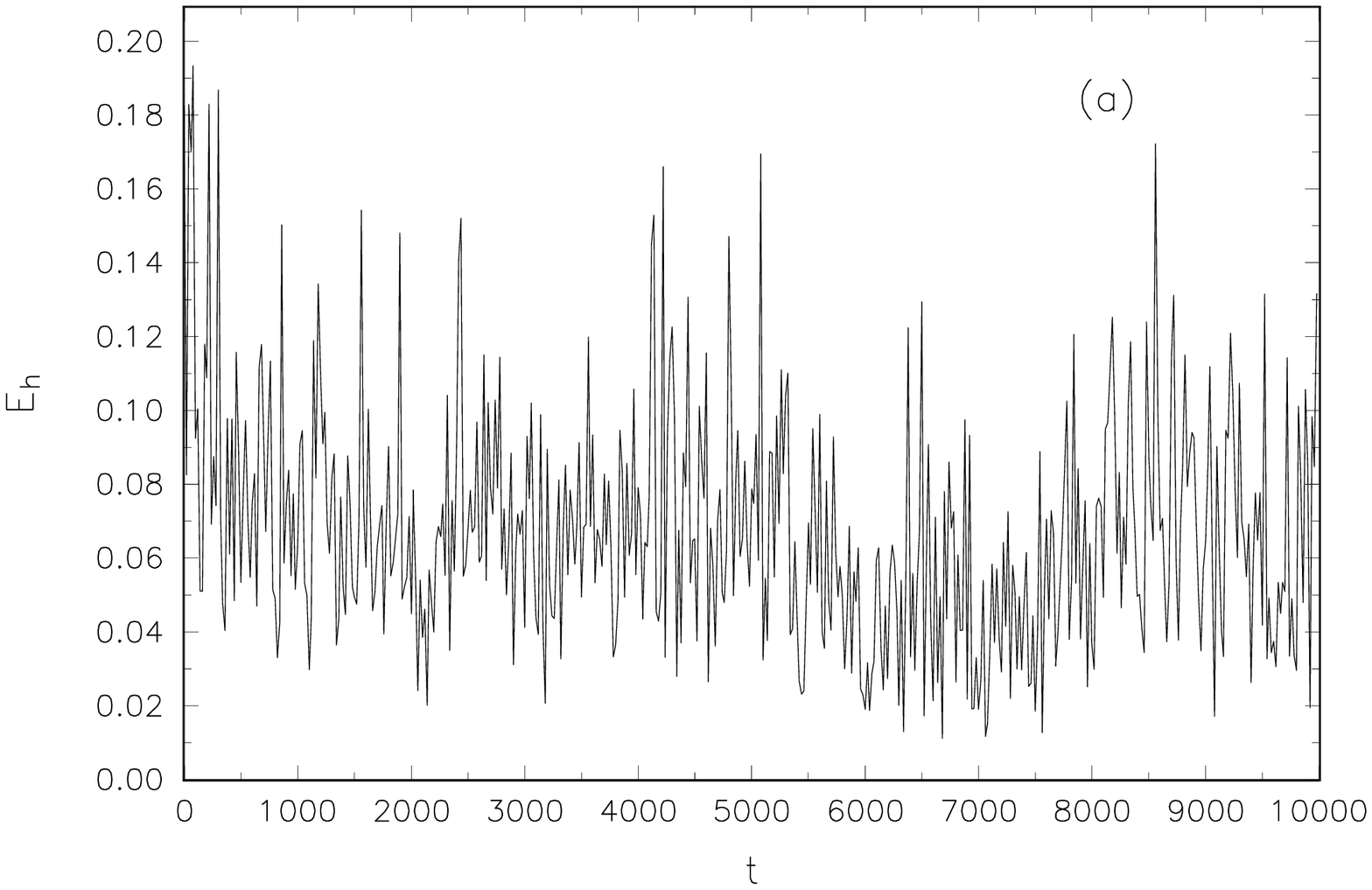}&
    \includegraphics[angle=-90,width=0.5\textwidth]{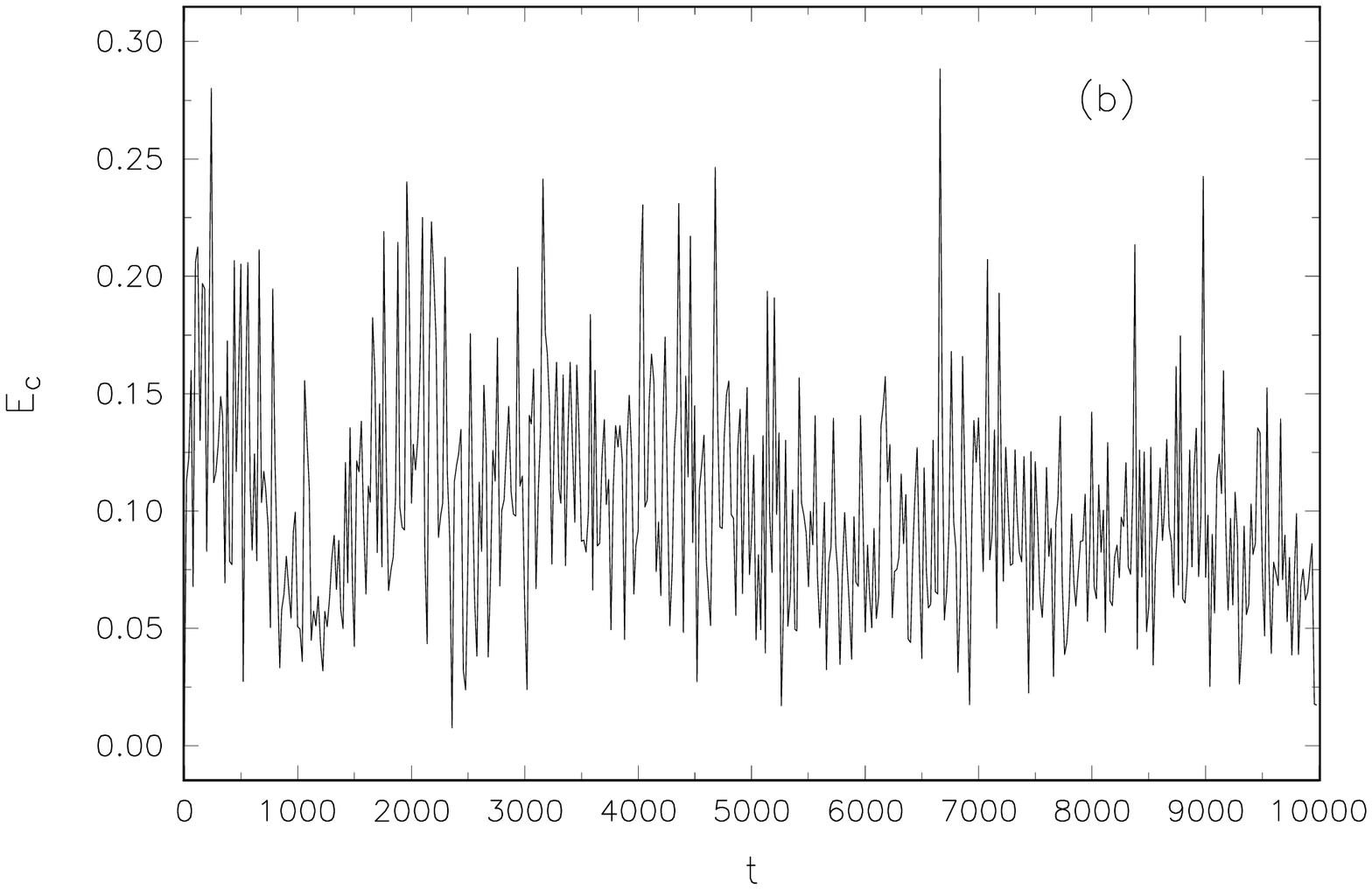}\\
    \includegraphics[angle=-90,width=0.5\textwidth]{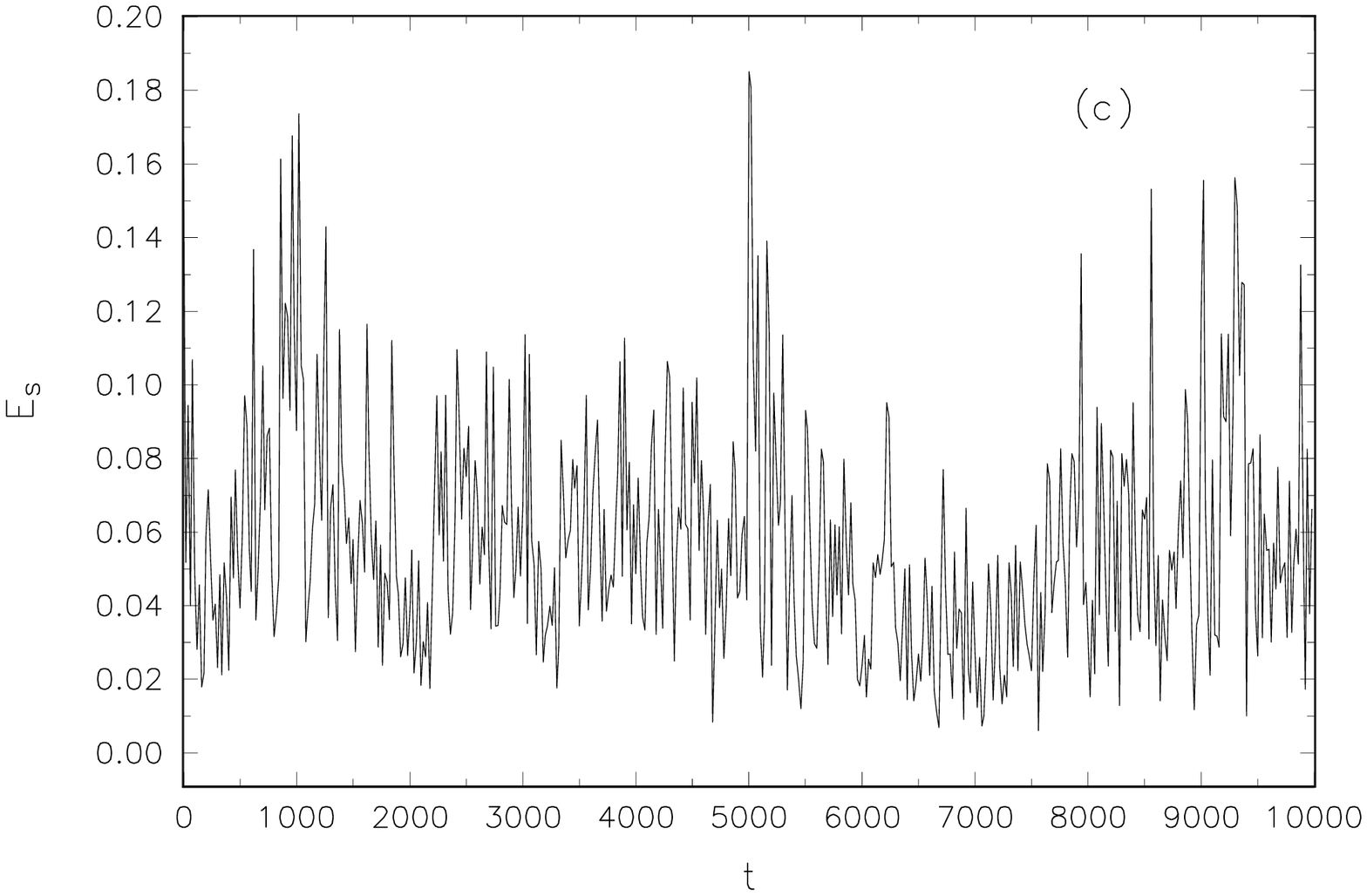}&
        \includegraphics[angle=-90,width=0.5\textwidth]{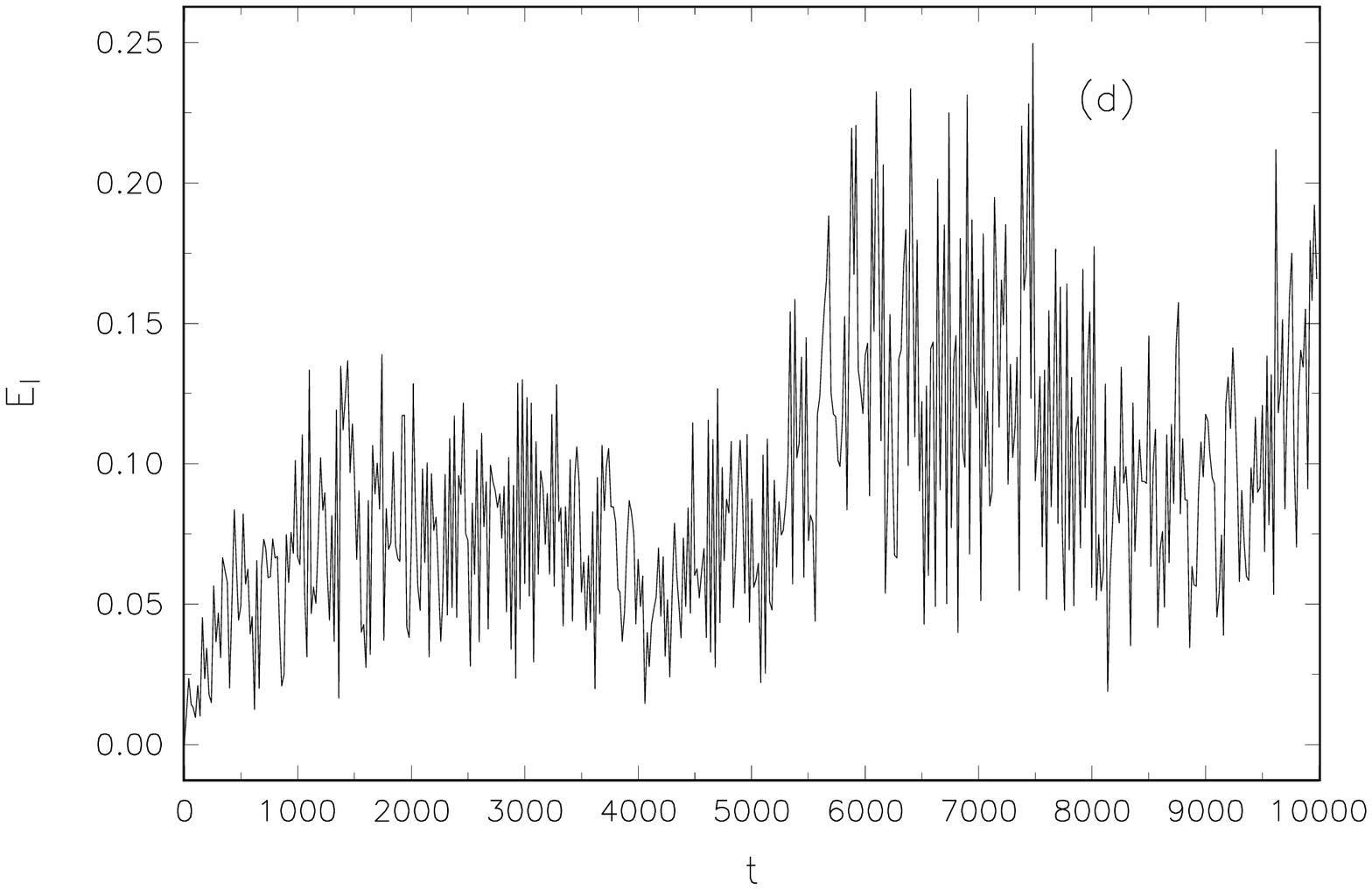}
\end{tabular}
  \caption{Relaxation dynamics for the twisted
system: Temporal behavior of the potential energies: (a) $V_h$,
(b) $V_c$. (c) $V_s$, and (d) $V_l$.
 } \label{fig:fig6}
\end{figure}

\subsection{Structural changes}
The dynamics of the structural changes in the course of the
relaxation process is displayed in Fig.\,$7$ and Fig.\,$8$ for the
ladder and twisted system, respectively. More precisely, we depict
the time evolution of the mean value of the lateral (radial)
deformations
\begin{equation}
\bar{d}(t)=\frac{1}{N}\,\sum_{n=1}^{N}\,d_n(t)\,,
\end{equation}
the angular deformations
\begin{equation}
\bar{\phi}(t)=\frac{1}{N}\,\sum_{n=1}^{N}\,(\theta_n(t)-(n-1)\,\theta_0)\,,
 \end{equation}
and the mean (axial) distance between two adjacent bases on a
strand is given by
\begin{equation}
\bar{l}(t)=\frac{1}{2N}\,\sum_{i=1,2}\,\sum_{n=1}^{N}\,l_{n,i}(t)\,,
\end{equation}
respectively. $\bar{d}$ ($\bar{l}$) determines also the average
length change of the hydrogen (covalent) bonds. We have directly
superposed the graphs from the pulling (with an inversed time) and
relaxation process. With perspective to the mean lateral
deformation $\bar{d}$ we observe that the relaxation process makes
fast progress relative to the pulling operation so that the mean
lateral deformation subsides rapidly. Conversely, the mean axial
deformation $\bar{l}$ shows immediately sizeable growth in
amplitude compared to the corresponding small changes during the
pulling process. Finally, for the twisted system the mean angular
deformations do not return to its starting zero value but remain
close to the level obtained at the end of the pulling procedure
such that the helix possesses reduced twist relative to the
equilibrium configuration.

\begin{figure}
  \begin{center}
  \begin{tabular}{cc}
    \includegraphics[angle=-90,width=0.5\textwidth]{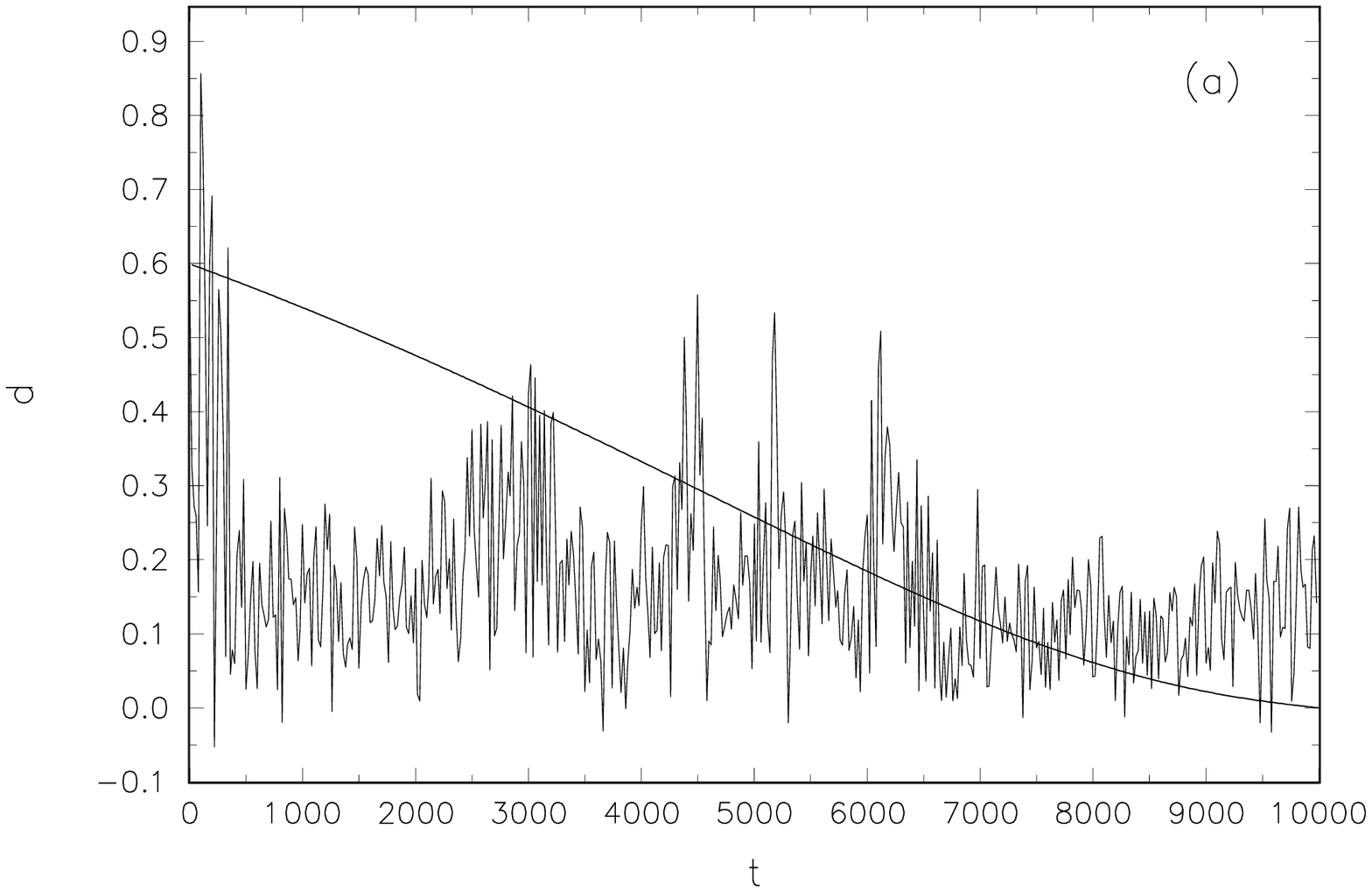}&
    \includegraphics[angle=-90,width=0.5\textwidth]{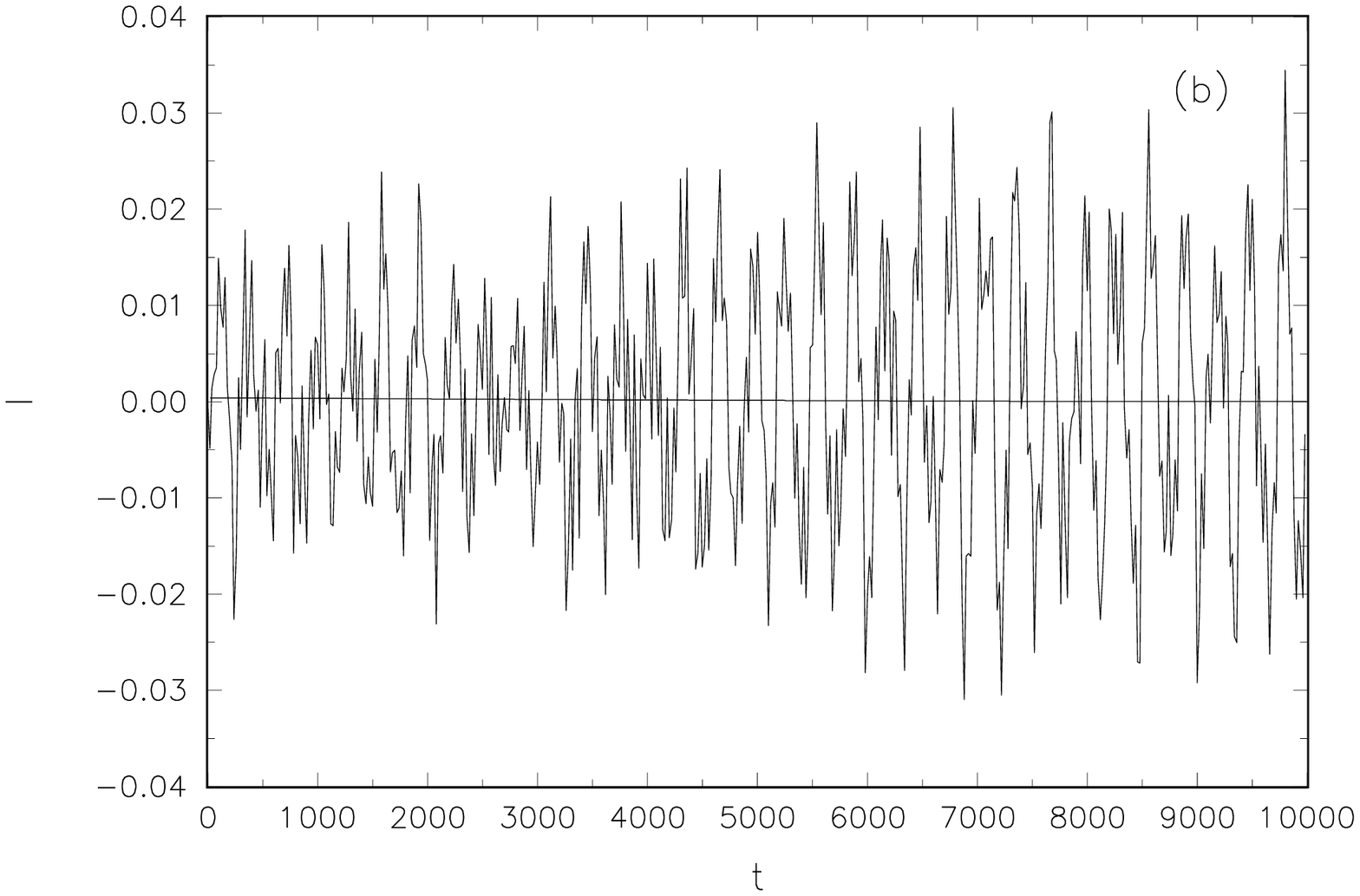}
  \end{tabular}
  \caption{Relaxation dynamics: Time-evolution of
the structural changes of the ladder system. (a) Mean lateral
deformations $\bar{d}(t)$. (b) Mean changes of the distances
between consecutive base planes $\bar{l}(t)$.
 } \label{fig:fig7}
\end{center}
\end{figure}

Comparing the mean lateral deformations of ladder and the twisted
system (starting in both cases from the value
$\bar{d}(0)=0.595\,\AA$) we conclude that for the twisted system
$\bar{d}(t)$ arrives by far much faster at the equilibrium mean
value $\bar{d}\sim 0.075\,\AA$ than it is the case for the ladder
system attaining the
 mean value $\bar{d}\sim 0.125\,\AA$. There remain
 permanent lateral
extensions of the ladder and the twisted system deviating from the
original equilibrium shape. On the other hand, the time evolution
of the corresponding variance, $\Delta
d=1/N\sum_{n=1}^{N}(d_n-\bar{d})^2$,
 reveals that the average
deviation from the mean value is markedly weaker in the twisted
case ($\Delta d \leq 0.032\,\AA$) compared to the considerably
strong deviations in the ladder case ($\Delta d \leq 0.084\,\AA$).
This shows that the attractive stacking interaction between
successive base pairs gets stronger in the twisted system favoring
the corresponding inter-base distances to retain faster a similar
value. In fact, those base pairs which have been forcefully
separated by the stretching recombine within less than $1000\,ns$
reversing the strand separation. Interestingly, the recombination
of the DNA duplex (associated mainly with the relaxation of the
lateral components) takes shorter times than the forced
separation. However, the relaxation is not complete because the
relaxed configuration with its fluctuating positions of the bases
still differs from the original equilibrium configuration with
resting base positions. Furthermore, distinct from the
non-fluctuating behavior during the pulling process the relaxation
curves exhibit fairly strong fluctuations. Evidently, the
relaxation (closing) does not take  place
  as the reversed process
of the pulling (opening).

Analogous to the temporal behavior of the lateral extensions the
mean distance between two successive bases on a strand
$\bar{l}(t)$ exhibits irregular oscillations such that for the
twisted system the large-amplitude fluctuations are up to two
times stronger than the ones for the ladder system. In contrast to
the permanently deformed lateral components (with non-zero mean
value $\bar{d}$) the average axial distance changes fluctuate
around the original equilibrium values $l_{n,i}=0$. Finally, for
the twisted case the mean twist angle performs small-amplitude
oscillations leaving the helix in slightly distorted angular
shape.
\begin{figure}[ht]
\begin{tabular}{cc}
    \includegraphics[angle=-90,width=0.5\textwidth]{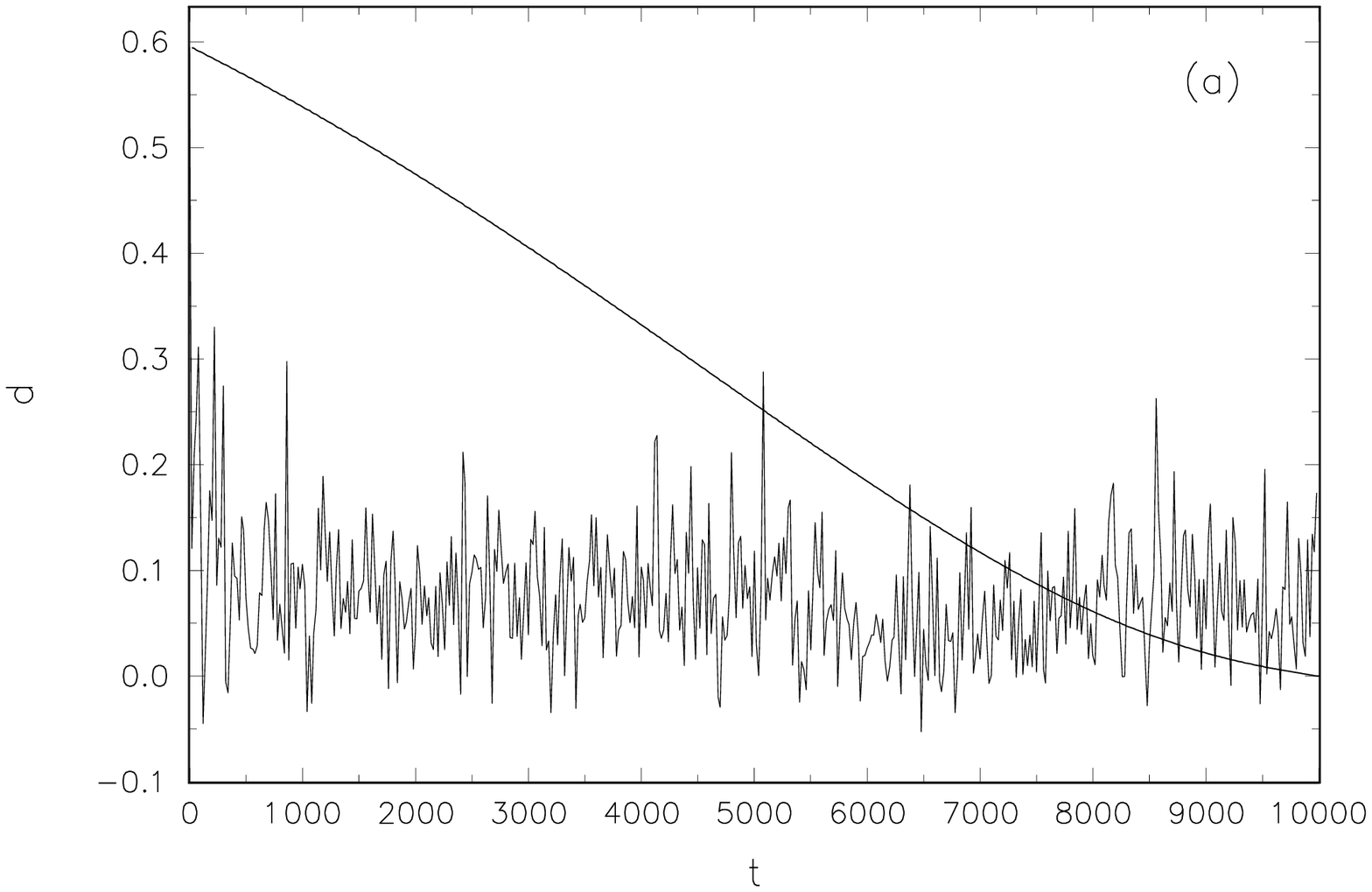}&
    \includegraphics[angle=-90,width=0.5\textwidth]{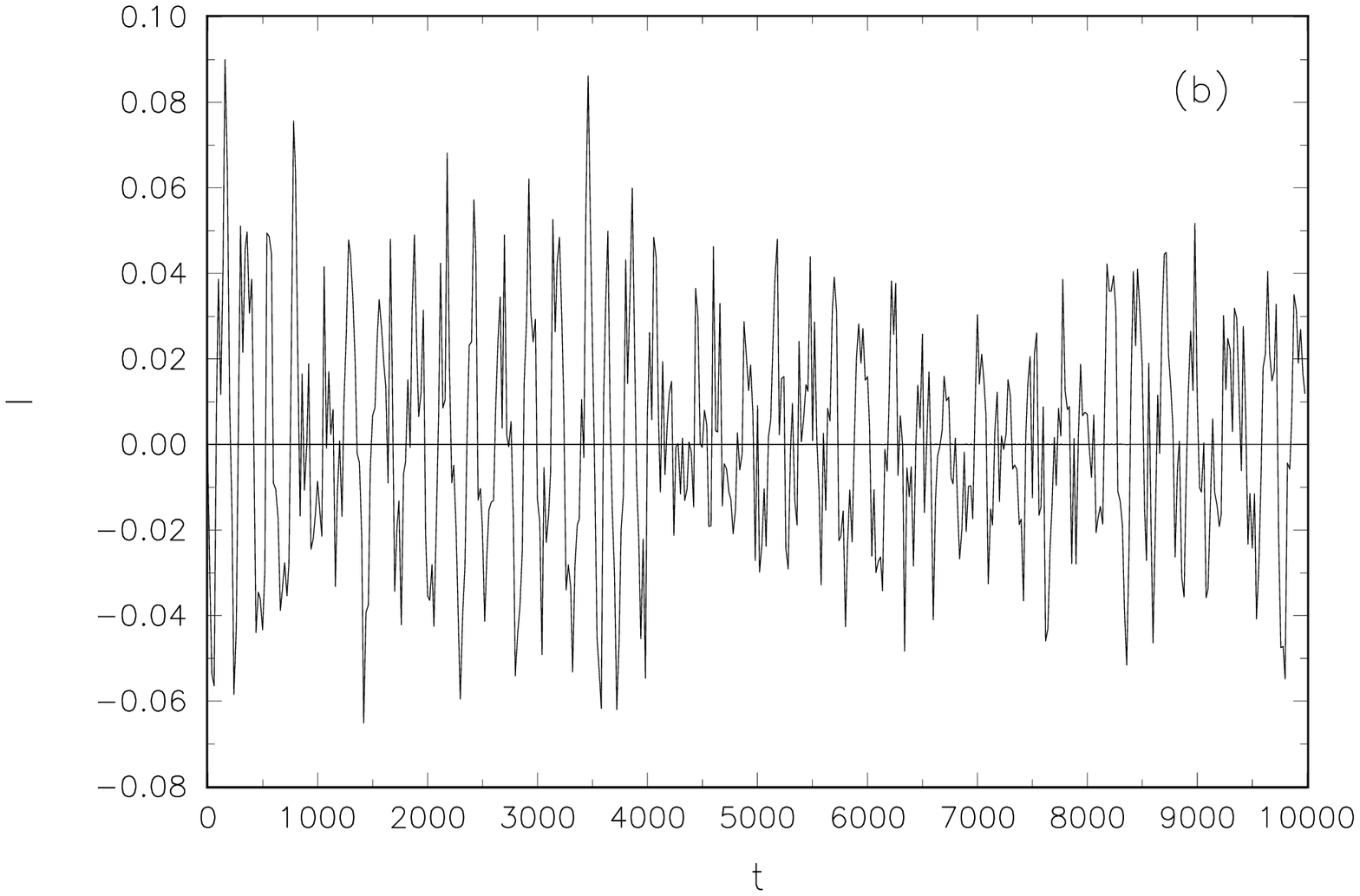}\\
\end{tabular}
\begin{center}
     \includegraphics[angle=-90,width=0.5\textwidth]{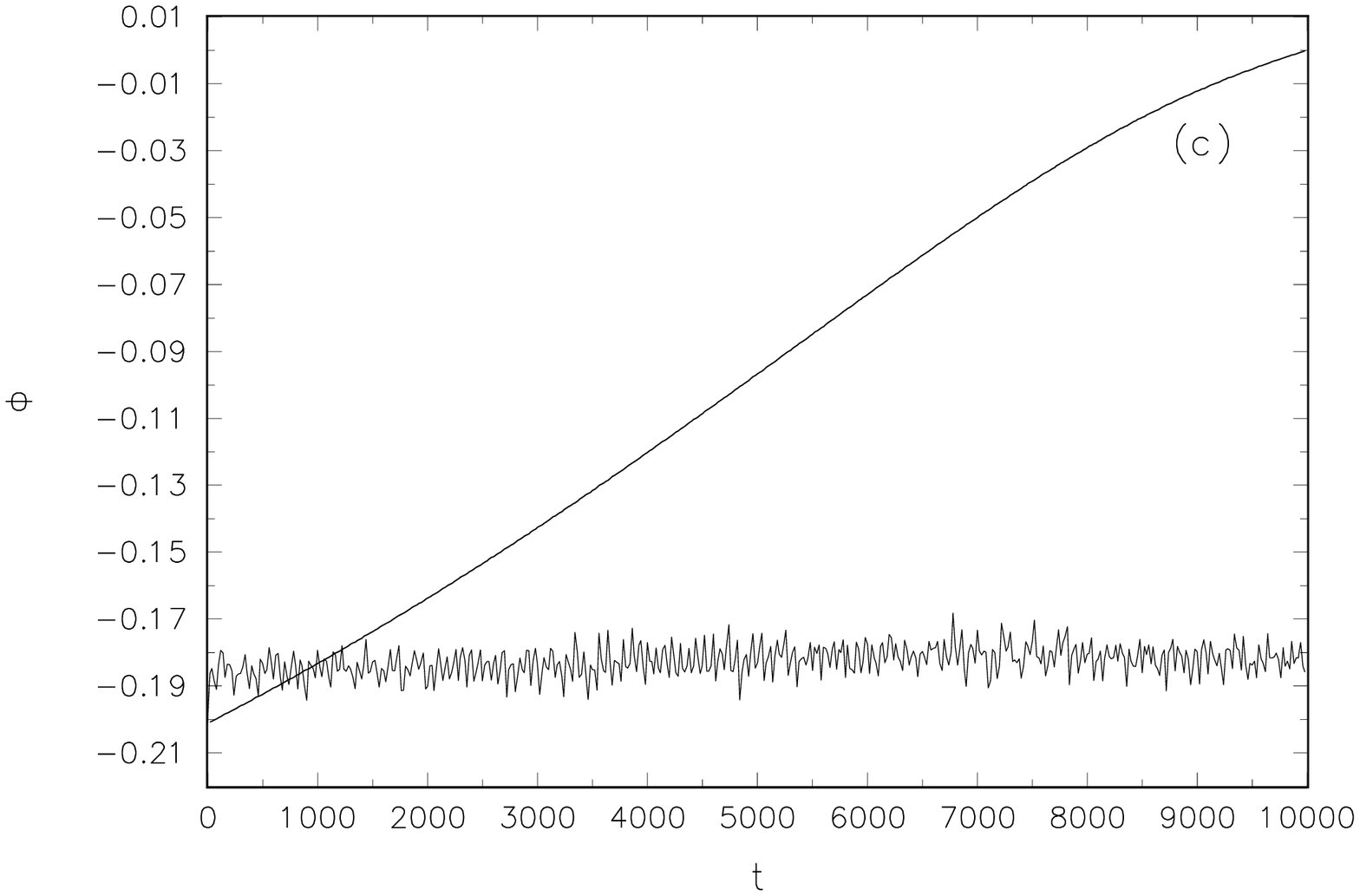}
  \caption{Relaxation dynamics: Time-evolution of
the structural changes of the twisted system. Superimposed is the
temporal behavior of the deformations during the pulling process
with an inversed time. (a) Mean lateral deformations $\bar{d}(t)$.
(b) Mean changes of the distances between consecutive base planes
$\bar{l}(t)$. (c) Mean angular deformation $\bar{\phi}(t)$.
 } \label{fig:fig8}
\end{center}
\end{figure}

\subsection{Comparison between the pulling and relaxation
processes}
 In summary, the ladder system and the twisted one have
in common that when both start from an initial partially opened
duplex corresponding to a lateral elongation pattern they reach
due to the relaxation process a quasi-equilibrium regime not far
from the initial equilibrium configuration. However, unlike for
the original equilibrium configuration in the newly adopted
quasi-equilibrium regime the structural coordinates are not at
rest but perform oscillatory motions with a mean period duration of
$50\,ns$. Correspondingly, this quasi-equilibrium regime is
characterized by energy exchanges between the various degrees of
freedom related with the alterations of the H-bridges, the
covalent bonds, the longitudinal elasticity potentials  and the
stacking interactions. Moreover, the adaptation process towards
the quasi-equilibrium takes much longer time in the ladder case
than in the twisted one. Hence the ladder system is more rigid
with respect to spatial deformations than its twisted counterpart
which with its additional angular degrees of freedom seems to
possess more flexibility but less adaptability. In particular we
observe that, with regard to the lateral deformations the twisted
double-stranded system almost reverts to the original
non-distorted equilibrium shape on the expense of substantial
fluctuations in axial direction. In contrast, the ladder system is
eventually left in comparatively strong lateral deformation but
without significant axial distance fluctuations pointing to strong
backbone rigidity. These results demonstrate that the relaxation
process
 does not take place
  as the complete reversal of the pulling
process. Similar behavior has been observed for the force-induced
melting of the DNA double helix and its subsequent reannealing
\cite{Clausen}.

Finally we remark that the velocity of the complete
recombination of the DNA
molecule is lower than the velocity imposed by the forcing unit
not uncommon for the opening-closing cycle of DNA molecules (see
e.g. \cite{Bockelmann},\cite{Thomen}).

\subsection{Guided relaxation}
 We have also  studied the elastic response of the
stretched DNA double chains
 for the twisted system
 when at the end of the pulling process the molecule
is not immediately exempted from the forcing unit.  The starting
point is an unopened molecule whose terminal end is pulled
laterally with a velocity $v_p=20\,\mu m s^{-1}$. When a certain
lateral extension of the terminal base pair is achieved the
direction of the pulling motion is reversed (without waiting time)
which returns the forcing unit to the position it had before
opening allowing for guided relaxation. We monitored
 the mean lateral extension $\bar{d}(t)$ for different values
of the velocity of the backward motion, namely $v_p=20\,\mu m
s^{-1}$, $v_p=10\,\mu m s^{-1}$ and $v_p=5\,\mu m s^{-1}$. The
results regarding the structural changes are shown in Fig.\,$9$
for backward velocity $v_b=20\,\mu m s^{-1}$. As expected, at the
end of the backward guidance of the terminal base pair the whole
of the double helix restores virtually the original (undistorted)
shape. Analogously, the associated deformation energies decrease
steadily in the course of the relaxation process and adopt their
initial zero content which they had before the pulling operation.
However, we observe an asymmetry between the forward and backward
elongation curves. Particularly, in an early stage of the
retraction of the pulling unit the corresponding relaxation curves
are in advance of their counterparts for the pulling process which
seems to present a  typical behavior of DNA molecules (see e.g.
\cite{Bockelmann1},\cite{Clausen},\cite{Cocco021})

With lowered backward velocity the process till complete
relaxation is achieved takes correspondingly longer times.

\begin{figure}[ht]
  \begin{center}
\begin{tabular}{cc}
    \includegraphics[angle=-90,width=0.5\textwidth]{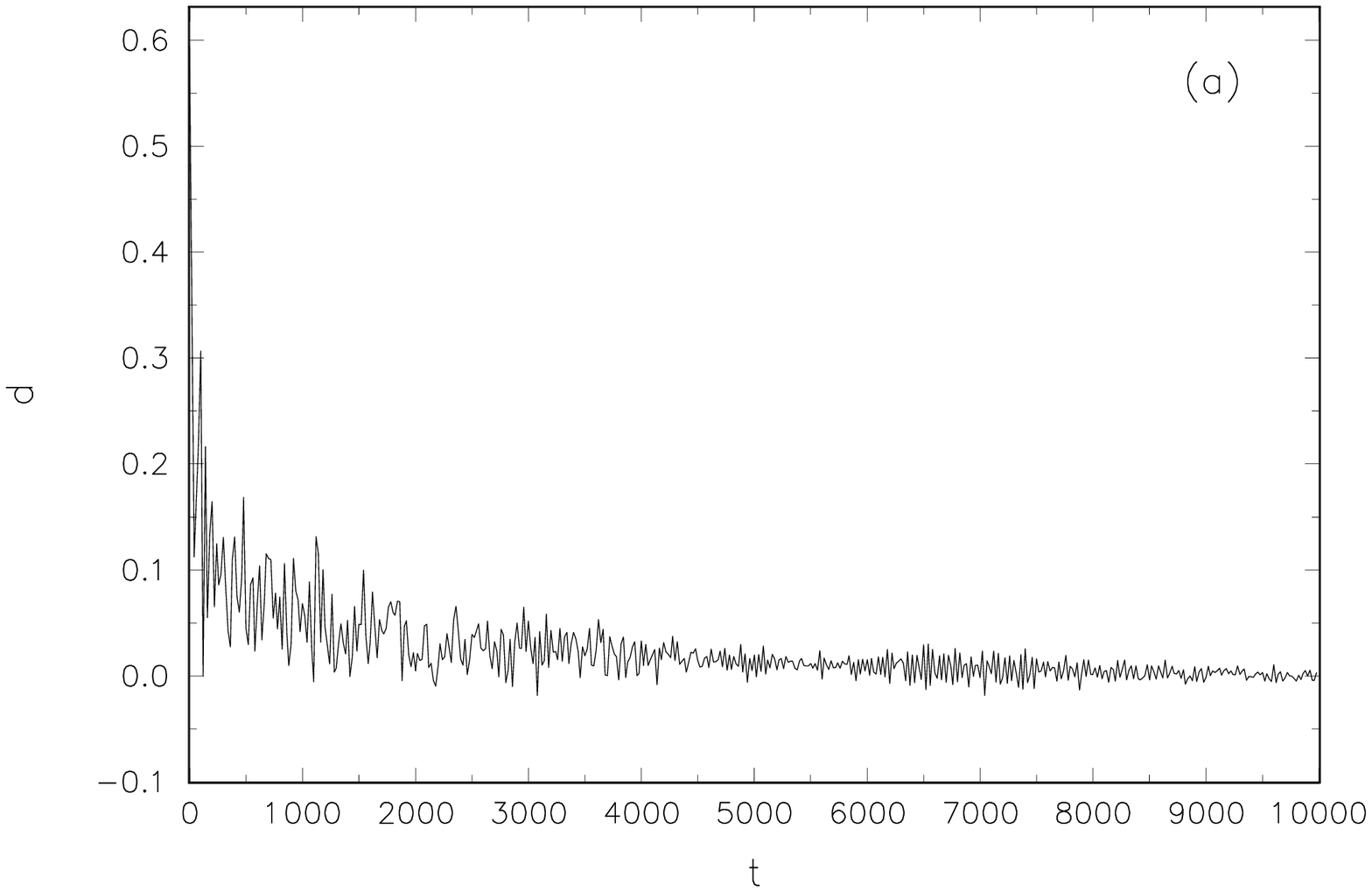}&
    \includegraphics[angle=-90,width=0.5\textwidth]{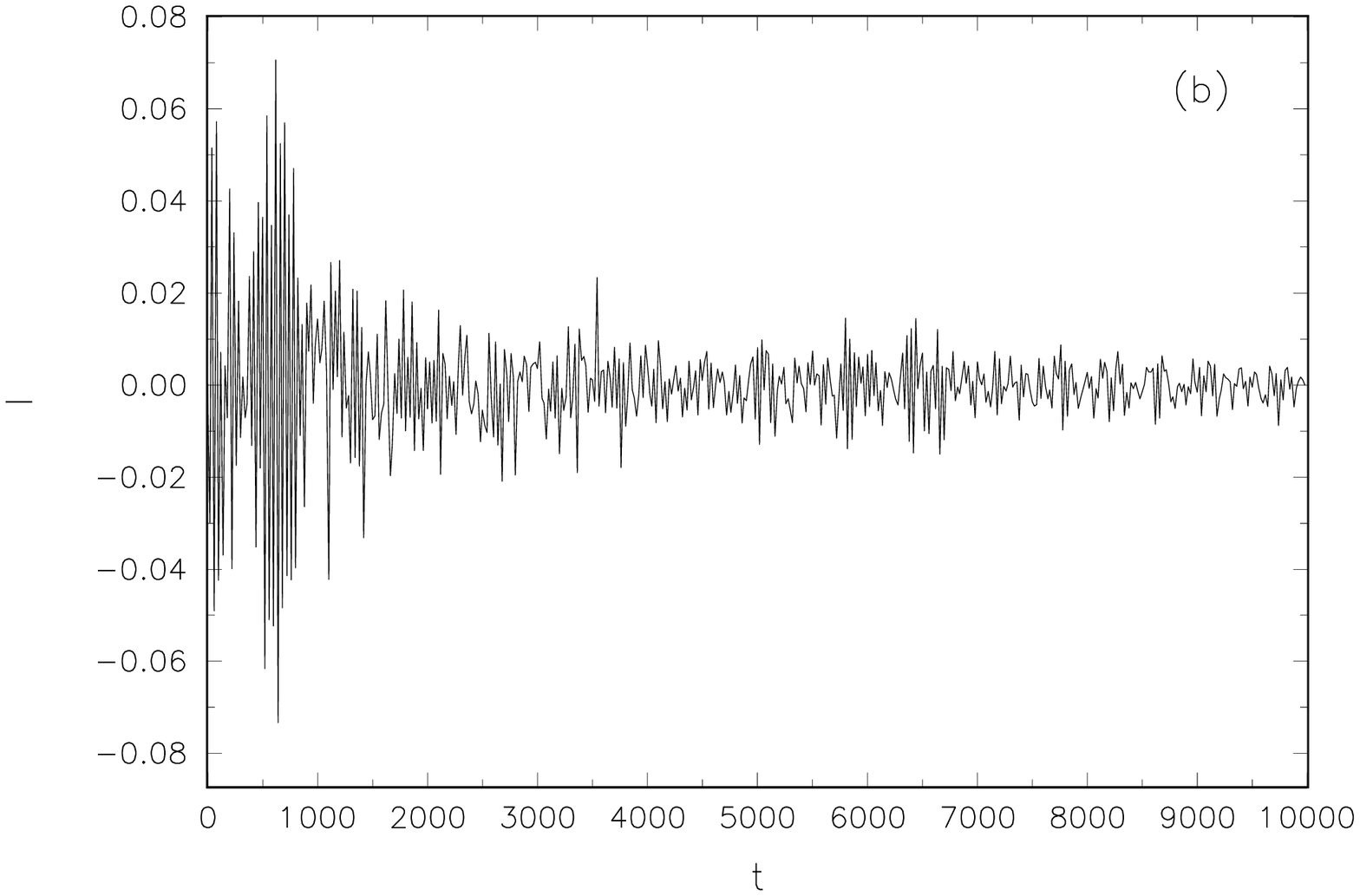}
\end{tabular}
    \includegraphics[angle=-90,width=0.5\textwidth]{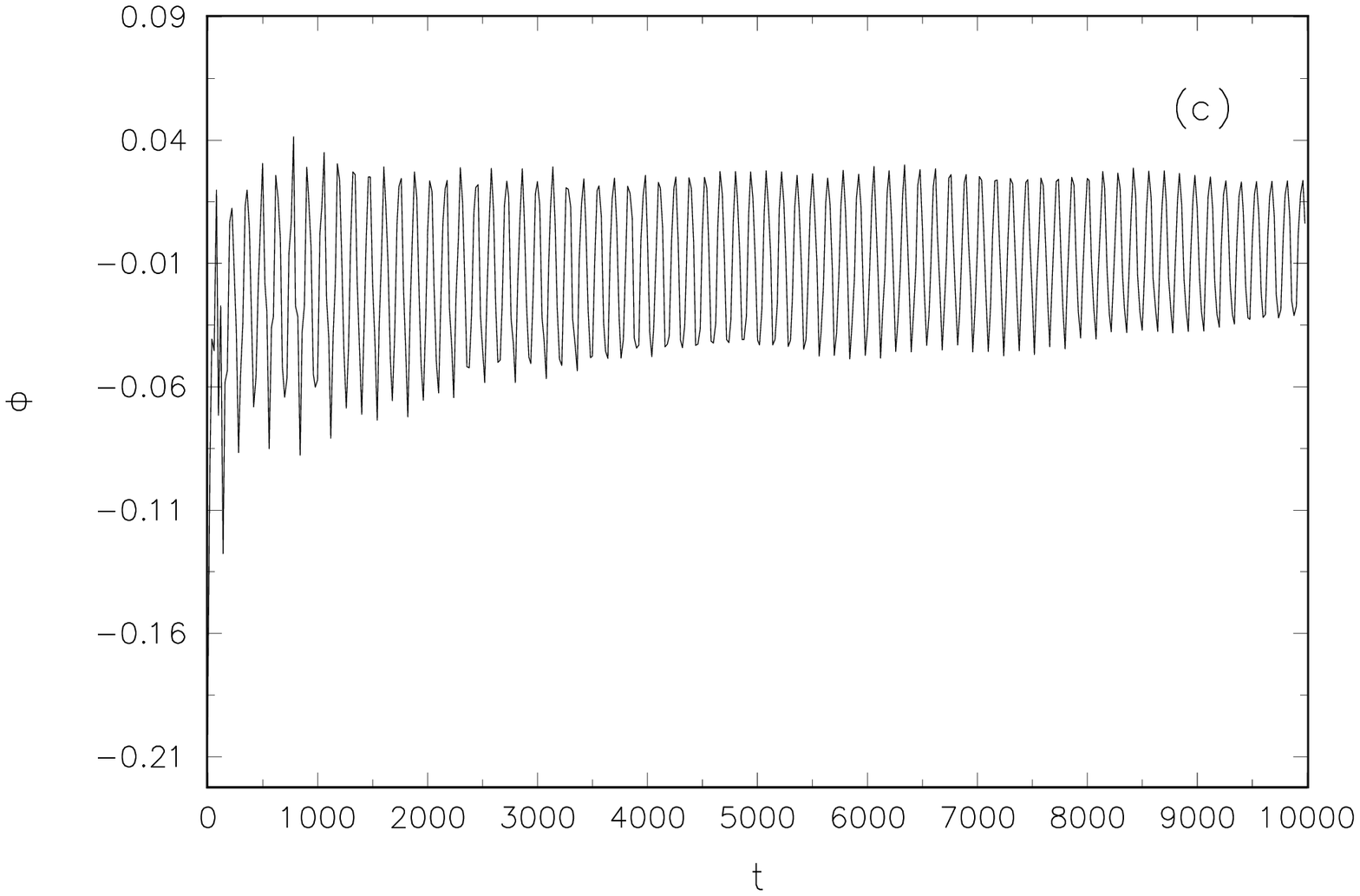}
  \caption{The relaxation dynamics for the twisted system for backward
velocity $v_b=20\mu m/s$. (a) Mean lateral deformations
$\bar{d}(t)$. (b) Mean changes of the distances between
consecutive base planes $\bar{l}(t)$. (c) Mean angular deformation
$\bar{\phi}(t)$.
 } \label{fig:fig9}
\end{center}
\end{figure}

\subsection{Parameter dependence}
For a parameter study we have varied the stacking interaction
strength $S$ and simulated the pulling process to follow
afterwards the relaxation dynamics. As the main result we found
that the larger $S$ is the more rapid proceeds the relaxation
process, viz. the faster is the quasi-equilibrium regime attained.
Furthermore, it holds in general that the energy migration takes
more time in the case of the ladder system than for its twisted
counterpart.

With view to a dependence on the molecule length $N$, i.e. the
number of base pairs, we remark that the time scale for the
relaxation process remains basically the same when $N$ is varied
in the range from $10$ to $30$ providing strong evidence for the
enormous elasticity of  DNA molecules on the lower amplitude
scale. In contrast, for large-amplitude elongations the
dissociation time of DNA molecules increases from a fraction of a
second to hours over the range of $N=10$ to $20$ base pairs
\cite{Cocco02}.

 In addition, with regard
to asymmetry we investigated also cases when the ends of the
strands are pulled laterally with unequal forces (in the extreme
case only one strand is pulled) resulting in elongation patterns
being asymmetric with respect to the central helix axis. We
observed then that the systems behave during  the subsequent
relaxation dynamics like in the symmetric cases discussed above
and attain finally quasi-equilibrium regimes being not far away
from the respective original equilibrium configurations with
fluctuating structural coordinates.

\section{Summary}\label{section:summary}

In this paper we have demonstrated that within the frame of an
oscillator network model we are able to study numerically the
mechanical stability and elasticity properties of DNA molecules.
Our model of coupled oscillators takes essential microscopic
degrees of freedom of DNA and the inherent interactions between
them into account. We have focused our interest on the
opening-closing dynamics of double-stranded DNA molecules. The
opening of a DNA molecule has been forced by mechanical stress
imposed at a terminal end of the molecule bringing it into a
non-equilibrium state. We have followed the subsequent relaxation
towards an equilibrium state with  recombination to its
double-stranded conformation. The similarities and differences
between the relaxation dynamics for a planar ladder-like
 DNA molecule and a twisted one have been
discussed. In particular we have shown that the attainment of a
quasi-equilibrium regime proceeds faster in the case of the
twisted DNA form than for its ladder counterpart. This has led us
to the conclusion that in the twisted form the DNA molecule is
more flexible with respect to conformational changes than in the
planar ladder-like version.

There are some limitations in our model approach and for an
improvement one should certainly consider the dependence on the
DNA sequence, the impact of the chemical
 environment
 and the influence of temperature on the mechanical stability.

\vspace{0.5cm}
\centerline{\large{\bf Acknowledgments}}

\noindent One of the authors acknowledges  support by the Deutsche
Forschungsgemeinschaft via a Heisenberg fellowship (He 3049/1-1).
The authors  would like to express their gratitude to the support
under the LOCNET EU network HPRN-CT-1999-00163.

\end{document}